\documentclass[aps,pra,superscriptaddress,twocolumn,floatfix,longbibliography]{revtex4-2}
\usepackage{dcolumn}
\usepackage{graphicx,color}
\usepackage[utf8]{inputenc}
\usepackage[T1]{fontenc}
\usepackage{amsmath,amssymb,bm,amsfonts,dsfont,mathrsfs,amsthm}
\usepackage{braket}
\usepackage{txfonts,comment}
\usepackage[colorlinks=true,linkcolor=blue,citecolor=blue,urlcolor=blue]{hyperref}
\usepackage{soul}
\usepackage{appendix}
\usepackage{float}
\usepackage{qcircuit}
\newcommand{\opi}[1]{\hat{#1}}

\begin{document}
\title{Probing the Limits of Dispersive Quantum Thermometry with a Nonlinear Mach-Zehnder-Based Quantum Simulator}

\author{Daniel Y. Akamatsu}
\affiliation{Instituto de F\'isica, Universidade Federal de Goi\'as, 74.001-970, Goi\^ania - GO, Brazil}
\thanks{These authors contributed equally to this work.}

\author{Lucas Ferreira R. de Moura}
\affiliation{Instituto de F\'isica, Universidade Federal de Goi\'as, 74.001-970, Goi\^ania - GO, Brazil}
\thanks{These authors contributed equally to this work.}

\author{Gabriella G. Damas}
\affiliation{Instituto de F\'isica, Universidade Federal de Goi\'as, 74.001-970, Goi\^ania - GO, Brazil}

\author{Gentil D. de Moraes Neto}
\affiliation{College of Physics and Engineering, Qufu Normal University, 273165 Qufu, China}
\email{gdmneto@gmail.com}

\author{Victor Montenegro}
\affiliation{College of Computing and Mathematical Sciences, Department of Applied Mathematics and Sciences, Khalifa University of Science and Technology, 127788 Abu Dhabi, United Arab Emirates}
\affiliation{Institute of Fundamental and Frontier Sciences, University of Electronic Science and Technology of China, Chengdu 611731, China}
\affiliation{Key Laboratory of Quantum Physics and Photonic Quantum Information, Ministry of Education, University of Electronic Science and Technology of China, Chengdu 611731, China}
\email{victor.montenegro@ku.ac.ae}

\author{Norton G. de Almeida}
\affiliation{Instituto de F\'isica, Universidade Federal de Goi\'as, 74.001-970, Goi\^ania - GO, Brazil}
\affiliation{Instituto de F\'isica de S\~ao Carlos, Universidade de S\~ao Paulo, Caixa Postal 969, 19560-970, S\~ao Carlos, S\~ao Paulo, Brazil}

\begin{abstract}
Temperature estimation, known as thermometry, is a critical sensing task for physical systems operating in the quantum regime. Indeed, thermal fluctuations can significantly degrade quantum coherence. Therefore, accurately determining the system's operating temperature is a crucial first step toward distinguishing thermal noise from other sources of decoherence. In this work, we estimate the unknown temperature of a collection of identical and independent two-level atoms dispersively probed by a single-mode quantized electromagnetic field. In contrast to previous works, we present an analytical sensing analysis demonstrating that the joint atom-field evolution---without any assumptions or approximations---can achieve, at best, the standard quantum limit of precision concerning the number of field excitations. To investigate our analysis further, we propose and implement a quantum thermometer based on a nonlinear Mach-Zehnder interferometer, which we realize through quantum digital simulation. Our simulation is highly flexible regarding atomic state preparation, allowing the initialization of atomic ensembles with positive and effective negative temperatures. This makes our platform a promising and versatile testbed for benchmarking thermometric capabilities in current quantum simulators.
\end{abstract}

\maketitle

\section{Introduction}

The rapid advancement of quantum technologies has enabled unprecedented control over small-scale systems at ultra-low temperatures~\cite{zoller2005quantum, giazotto2006opportunities, pekola2015towards,timofeev2009electronic}. This progress has driven both the growth of quantum thermodynamics as a field~\cite{kosloff2013quantum, goold2016d, sai2016quantum, millen2015perspective, deffner2019quantum} and the development of metrological strategies for high-precision temperature estimation, known as thermometry~\cite{mohammady2018low, mehboudi2019thermometry,de2018quantum,purdy2015optomechanical, montenegro2020mechanical,razavian2019quantum,mihailescu2023thermometry,srivastava2023topological,plodzien2018few,sekatski2022optimal, correa2015individual, mukherjee2019enhanced, glatthard2022optimal,mok2021optimal,rubio2021global,campbell2017global,alves2022bayesian,jorgensen2022bayesian,mihailescu2023multiparameter, ostermann2023temperature, aybar2022criticalquantum,asghar2023low,mehboudi2022fundamental,campbell2018precision}. Indeed, thermometry plays a vital role across a broad range of fields~\cite{quintanilla2022challenges, brites2012thermometry, kucsko2013nanometre, fujiwara2021diamond, dacanin2023luminescence, chu2022thermodynamic}. In particular, understanding thermal effects in quantum systems is crucial~\cite{efetov1995temperature, lopes2020finite, watanabe2022quantum, lucrezi2024temperature, ahumada2023embedded, xu2024persisting}, especially for Noisy Intermediate-Scale Quantum (NISQ) devices, where thermal noise can severely degrade quantum coherence~\cite{bharti2022noisy}. Hence, achieving highly accurate thermometry is essential.

To address this task, quantum parameter estimation theory~\cite{helstrom1969quantum,paris2009quantum} provides a rigorous framework to evaluate how well a quantum probe can estimate unknown parameters---such as temperature~\cite{mehboudi2019thermometry,de2018quantum}---in an operational and quantitative way. Within this framework, the fundamental bound on precision is characterized by the quantum Fisher information (QFI)~\cite{helstrom1969quantum, vantrees1968detection}, a Riemannian metric that quantifies the distinguishability between a quantum state and its infinitesimally neighboring state~\cite{braunstein1994statistical}. The QFI sets the ultimate precision limit with which a specific quantum probe can estimate a parameter, given a fixed sensing resource---such as total time~\cite{yuan2025utilizing,montenegro2022sequential,yang2023extractable,albarelli2018restoring,rossi2020noisy,smirne2016ultimate,montenegro2023quantum}, number of particles~\cite{giovannetti2004quantum, giovannetti2006quantum, giovannetti2011advances,montenegro2025quantum}, or available energy~\cite{cavazzoni2025frequency, montenegro2025enhanced}. For thermometry, when the probe is typically assumed to be in thermal equilibrium, the QFI reduces to the energy fluctuations of the probe, thereby constraining the achievable precision to the probe's heat capacity~\cite{jahnke2011operational,mehboudi2019thermometry,de2018quantum,correa2015individual}. To overcome this limitation, various dynamical approaches~\cite{mukherjee2019enhanced,campbell2017global,purdy2015optomechanical,oconnor2024fisher,razavian2019quantum,montenegro2020mechanical,yang2024quantum} have been proposed. In particular, it has been shown that in a dispersive atom-light system, the uncertainty in temperature estimation can decrease proportionally to the inverse square of the number of field excitations~\cite{stace2010quantum}---a sensing scaling known as the Heisenberg limit. However, while valid under specific conditions, this result relies on particular assumptions~\cite{stace2010quantum}. This naturally leads to two key questions: (i) Are these assumptions generally applicable? and (ii) Can they be implemented in realistic physical systems? In the following, we examine both questions in detail.

In this work, we estimate the unknown temperature encoded in the dynamical quantum state of a dispersive atom-light system. Specifically, we consider an ensemble of identical and independent two-level atoms at an unknown temperature $T$, coupled dispersively to a single-mode quantized electromagnetic field. Once the interaction is switched on, the temperature information becomes imprinted in the resulting correlated atom-field state. We analyze the ultimate precision limits set by the QFI, both for the entire system and when measurements are restricted to the light degree of freedom only. We analytically demonstrate that, even when the field is initialized in a maximally correlated state, the ultimate thermometric precision is fundamentally limited to the standard quantum limit---in stark contrast to previous claims. Importantly, our results are exact and do not rely on any approximations or assumptions about the system dynamics.

To further support and extend our findings, we propose an interferometric scheme that not only illustrates our main results, but also offers a feasible implementation using currently available superconducting quantum platforms. Indeed, among quantum-enhanced techniques, interferometric methods have attracted considerable interest due to their ability to achieve high precision through quantum-enhanced phase estimation~\cite{yang2024quantum, giovannetti2004quantum,giovannetti2006quantum,giovannetti2011advances,montenegro2025quantum,degen2017quantum}. In particular, the Mach-Zehnder interferometer (MZI) has been proposed as a powerful tool for temperature sensing, harnessing quantum superposition to amplify small thermal fluctuations~\cite{wilkie1963measurement, yadav2024quantum, grond2011mach, stace2010quantum}. To this end, we propose a nonlinear MZI for thermometric analysis, where the nonlinearity accounts for the generation of strongly nonclassical states---such as N00N states---in the field mode. We implement this scheme on a digital quantum simulator~\cite{qiskit,ibmq}, where the standard quantum limit of precision is achieved experimentally using single-photon states, and verified through simulations for higher photon number states. Furthermore, we demonstrate the flexibility of our approach by applying it to atomic ensembles prepared at both positive and effective negative temperatures. The latter, being highly relevant both theoretically~\cite{de2019efficiency,damas2023negative} and practically~\cite{purcell1951nuclear,ramsey1956thermodynamics,xi2017quantum, lincoln2013negative,yan2025anoff, braun2013negative,mendonca2020reservoir}.

\section{Quantum estimation background}

Quantum estimation theory (QET) provides a rigorous framework for determining the fundamental precision limits in estimating an unknown parameter $\theta$ encoded in a quantum state $\rho(\theta)$~\cite{helstrom1969quantum,paris2009quantum}. For simplicity, let us consider a single unknown parameter to estimate, while the remaining parameters are known and tunable at will. In this scenario, QET establishes the fundamental lower bound for estimating $\theta$ via the quantum Cram\'{e}r-Rao theorem~\cite{paris2009quantum, helstrom1969quantum,cramer1999mathematical,LeCam-1986,Holevo}.
\begin{equation}
    \mathrm{Var}[\check{\vartheta}] \geq \frac{1}{\nu \mathcal{F}(\theta)}\geq \frac{1}{\nu Q(\theta)},~\label{eq_quantumCRB}
\end{equation}
where $\mathrm{Var}[\check{\vartheta}]$ is the variance of the (assumed unbiased) local estimator $\check{\vartheta}$, $\check{\vartheta}$ is a function that maps all measurement outcomes to an estimate in parameter space, $\nu$ is the total number of experimental trials, and $\mathcal{F}(\theta)$ [$Q(\theta)$] is the classical (quantum) Fisher information function~\cite{paris2009quantum}. The classical Fisher information (CFI) is defined as (assuming discrete measurement outcomes):
\begin{equation}
    \mathcal{F}(\theta) = \sum_{k} \frac{1}{p_k(\theta)}\left[ \frac{\partial}{\partial\theta}p_k(\theta)\right]^2.\label{eq_CFI}
\end{equation}
In the above, $p_k(\theta) = \mathrm{Tr}[\hat{\Pi}_k \rho(\theta)]$ is the probability distribution for the measurement outcome $k$ given a value of $\theta$, where $\hat{\Pi}_k$ is a positive operator-valued measure (POVM). Intuitively, the CFI in Eq.~\eqref{eq_CFI} increases as the slope of $p_k(\theta)$ with respect to $\theta$ becomes steeper. This implies that more information about $\theta$ is gained when the probability distributions change more abruptly with $\theta$. Note that the CFI in Eq.~\eqref{eq_CFI} depends on the choice of POVM (i.e., its associated probability distributions). The optimal POVM that maximizes the CFI defines the QFI: 
\begin{equation}
Q(\theta) = \max_{\{\hat{\Pi}_k\}} \mathcal{F}(\theta).
\end{equation}
Thus, after maximizing over all possible POVMs $\{\hat{\Pi}_k\}$, the QFI sets the ultimate sensing capability of the quantum probe. Interestingly, geometric aspects of QET identify the QFI as the function that quantifies the distance between neighboring states as $\theta$ varies. Hence, higher sensing precision is achieved for states that are farther apart in the infinitesimal vicinity of $\theta$~\cite{sidhu2020geometric, braunstein1994statistical, venuti2007quantum, zanardi2006ground}.

In particular, the QFI for a density matrix expressed in spectral decomposition, $\rho(\theta) = \sum_j \lambda_j |\lambda_j\rangle\langle \lambda_j|$, where $\lambda_j$ is the eigenvalue of $\rho(\theta)$ and $|\lambda_j\rangle$ is its eigenvector, is given by~\cite{paris2009quantum}:
\begin{equation}
    Q(\theta) = 2 \sum_{j,k} (\lambda_j + \lambda_k)^{-1}| \langle \lambda_j | \frac{\partial}{\partial\theta} \rho(\theta) | \lambda_k \rangle |^2.\label{eq_QFI}
\end{equation}
Since we are focusing on thermometry limits, Eq.~\eqref{eq_QFI} will be sufficient for the purposes of this work.

\subsection{Thermometry preliminaries}

The quantum state of a system in thermal equilibrium at temperature $T$ is given by the Gibbs state (Boltzmann constant set as $k_B=1$):
\begin{equation}
    \rho(T) = \frac{e^{-\frac{1}{T}\hat{H}}}{Z(T)},\label{eq_Gibbs}
\end{equation}
with $\hat{H}$ being the Hamiltonian of the system and $Z(T)=\mathrm{Tr}[e^{-\frac{1}{T}\hat{H}}]$ is the partition function. The goal of thermometry is to estimate the unknown temperature $T$. By substituting the Gibbs state from Eq.~\eqref{eq_Gibbs} into Eq.~\eqref{eq_QFI}, one can straightforwardly evaluate the QFI as~\cite{mehboudi2019thermometry}:
\begin{equation}
    Q(T)=\frac{1}{T^4}(\Delta \hat{H})^2,
\end{equation}
where $(\Delta \hat{H})^2{=}\langle \hat{H}^2\rangle{-}\langle \hat{H} \rangle^2$ is the energy fluctuations. The latter expression highlights a clear limitation of thermometry at equilibrium, namely: to improve thermometry precision, it is necessary to increase the probe's energy fluctuations---or, equivalently, to increase the probe's heat capacitance $C_T = \frac{1}{T^2}(\Delta \hat{H})^2$~\cite{mehboudi2019thermometry}. This limitation can be relaxed by dynamically encoding the temperature in a joint system~\cite{stace2010quantum}. Indeed, this is the sensing strategy we will pursue throughout this work: a dynamical encoding of the temperature into an atom-field joint system.

\section{The theoretical model}

Consider a sample of $M$ non-interacting identical two-level atoms (qubits) with energy splitting $\varepsilon$ at thermal equilibrium with unknown temperature $T$. The state is described by the Gibbs state of Eq.~\eqref{eq_Gibbs}
\begin{equation}
    \rho_\mathrm{qubits}(T) = \frac{e^{-\frac{1}{T}\hat{H}_\mathrm{qubits}}}{Z(T)},\label{eq_Gibbs_qubits}
\end{equation}
with the Hamiltonian given by:
\begin{equation}
\hat{H}_\mathrm{qubits} = \varepsilon\sum_{k=1}^{M} \hat{\sigma}_+^{(k)}\hat{\sigma}_-^{(k)} = \frac{\varepsilon}{2} \sum_{k=1}^{M} (\hat{\sigma}_z^{(k)} + \hat{I}^{(k)}),\label{eq_Hqubits}
\end{equation}
where $\hat{\sigma}_+^{(k)} = (|e\rangle\langle g|)^{(k)}$, $\hat{\sigma}_-^{(k)} = (|g\rangle\langle e|)^{(k)}$, and $\hat{\sigma}_z^{(k)} = (|e\rangle\langle e| - |g\rangle\langle g|)^{(k)}$ are the Pauli operators acting at site $k$, and $\hat{I}^{(k)}=(|e\rangle\langle e| + |g\rangle\langle g|)^{(k)}$ is the identity matrix for the $k$th two-level atom. The corresponding QFI for the quantum state in Eq.~\eqref{eq_Gibbs_qubits} is given by:
\begin{equation}
    Q(T)=M \frac{\varepsilon^2}{4T^4} \mathrm{sech}^2\left(\frac{\varepsilon}{2T}\right).\label{qfi_single}
\end{equation}
Note that the linear scaling with $M$ is a direct consequence of the sample consisting of $M$ non-interacting two-level atoms.

Although Eq.~\eqref{eq_QFI} and Eq.~\eqref{qfi_single} provide closed-form expressions for the QFI functions, the optimal measurements required to reach the ultimate thermometry limit (i.e., energy measurements~\cite{jahnke2011operational, stace2010quantum}) are not always feasible~\cite{yang2024sequential}. See Refs.~\cite{hovhannisyan2021optimal, oconnor2024fisher} for coarse-grained versions of energy measurements and Ref.~\cite{glatthard2023energy} for thermometry beyond standard open-system weak-coupling assumptions. Furthermore, as shown in Eq.~\eqref{eq_QFI}, systems in thermal equilibrium limit the flexibility of the probe~\cite{mehboudi2019thermometry}. To overcome these challenges, we use a dynamical sensing strategy in which the temperature of the $M$ two-level atoms is probed indirectly through two bosonic field modes, which are assumed to be fully accessible and allow for feasible measurements.
\begin{figure}[t]
\centering
\includegraphics[width=\linewidth]{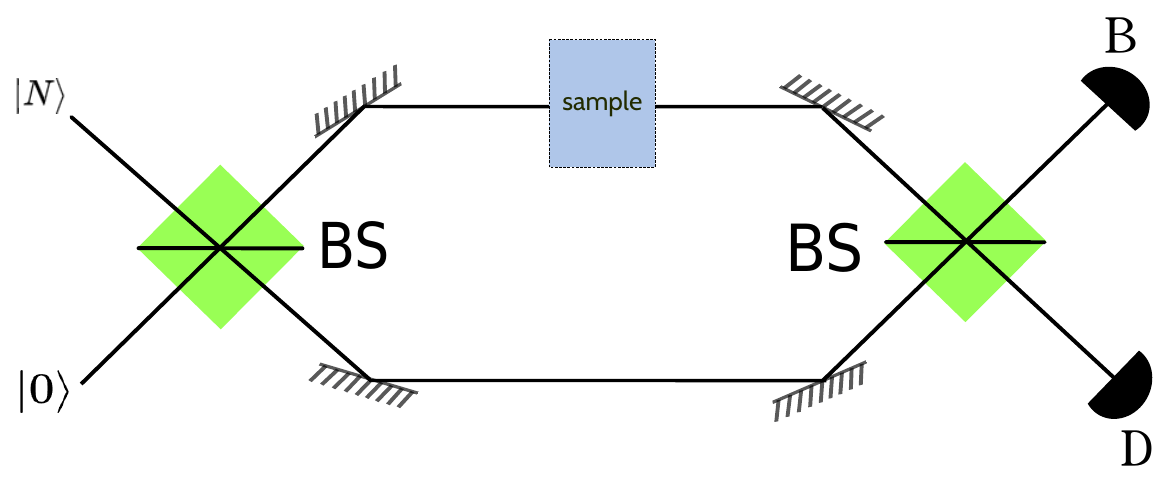}
\caption{Sketch of the Mach-Zehnder (MZ) thermometry scheme: A two-mode maximally entangled N00N state is generated by sending an initial separable input state $|N, 0\rangle{=}(|N{,}0\rangle{+}|0, N\rangle)/\sqrt{2}$ through a nonlinear beam splitter (BS). One mode of a N00N state interacts with a sample of $M$ identical, non-interacting two-level atoms (qubits) for a time $t$, with interaction strength $\chi$. The qubits are in thermal equilibrium at an unknown temperature $T$. The goal is to estimate this temperature $T$. After the interaction, the two modes are recombined at a second BS. Finally, a measurement is performed at the bright port (B) to estimate the unknown temperature $T$. The dark port is denoted as (D).}
\label{fig_sketch_MZ}
\end{figure}

In particular, we propose probing the sample using a MZI based on the Elitzur-Vaidman scheme~\cite{Elitzur93}, see Fig.~\ref{fig_sketch_MZ}. In the described MZI setup, if no object is placed in any arm, every incident photon is directed toward a designated detector—termed the \emph{bright port}—while the complementary detector is referred to as the \emph{dark port}. In contrast, when an object is introduced in one of the interferometer arms, the probability distribution of the detection events is altered. Notably, a photon detection at the dark port signals the presence of the object, even though the photon has not interacted with it directly. This interaction-free measurement can be achieved with a probability arbitrarily close to unity \cite{kwiat1995interaction}. Note, however, that in the original Elitzur-Vaidman proposal the object is completely absorbing, unlike our scheme in which the interaction is dispersive and the photon is never absorbed. However, even the dispersive interaction modifies the phase in the MZI causing the dark detector to start counting photons.

We probe the sample of $M$ non-interacting two-level atoms using a field mode under a dispersive Hamiltonian:
\begin{equation}
    \hat{H}=\varepsilon\chi \hat{a}^\dagger \hat{a} \hat{M}_\mathrm{qubits}, \hspace{1cm} \hat{M}_\mathrm{qubits}\equiv\sum_{k=1}^M \hat{\sigma}_+^{(k)}\hat{\sigma}_-^{(k)},\label{eq_H_interaction}
\end{equation}
where $\hat{M}_\mathrm{qubits}$ counts the total qubits' excitations, $\chi$ is a known coupling strength, $\hat{a}$ ($\hat{a}^\dagger$) is the bosonic annihilation (creation) operator for the field mode---obeying $ [\hat{a}, \hat{a}^\dagger] = 1 $---and $ \hat{a}^\dagger \hat{a} |n\rangle = n |n\rangle $, with $ |n\rangle $ being a Fock state containing $ n $ excitations.

We determine the dynamical thermometry limits under the assumption that only the single field $a$ mode is accessible, namely:
\begin{equation}
\rho_a(t)=\mathrm{Tr}_{\backslash a}\left[\hat{U}(t)\rho_0\hat{U}^\dagger(t)\right],
\end{equation}
where $\mathrm{Tr}_{\backslash a}$ corresponds to tracing out everything except party $a$, $\hat{U}(t) = e^{-it\hat{H}}$ (in Planck units $\hbar = 1$), and
\begin{equation}
    \rho_0 = \sum_{n,m,n',m'}\rho_{n,m,n',m'}|n,m\rangle\langle n',m'|\otimes\rho_\mathrm{qubits}(T).\label{eq_initial_state}
\end{equation}
Note that the above state $|n,m\rangle := |n\rangle_a \otimes |m\rangle_b$ represents the \textit{photon} number states of modes $a$ and $b$ in the MZI. Here, only mode $a$ interacts with the sample $\rho_\mathrm{qubits}(T)$, while mode $b$ propagates freely through the \textit{lower} arm of the interferometer, see Fig.~\ref{fig_sketch_MZ}. Finally, we compute the QFI with respect to $T$, using the quantum state $\rho_a(t)$.

\subsection{N00N input field}

The maximally entangled two-mode N00N state,
\begin{equation}
    |\text{N00N}\rangle = \frac{1}{\sqrt{2}} \left( |\text{N,0}\rangle + |\text{0,N}\rangle \right),
\end{equation}
is known to enable Heisenberg-limited precision in various sensing contexts~\cite{giovannetti2004quantum, giovannetti2006quantum, giovannetti2011advances}. This means that the QFI with respect to an unknown parameter scales as $N^2$, where $N$ is the number of (N00N) field excitations. In addition, it has been claimed that under strict conditions, quantum thermometry can also reach this limit~\cite{stace2010quantum,de2018quantum}. The following derivation, carried out without imposing additional constraints, establishes that quantum thermometry based on dispersive interactions is inherently bounded by the standard quantum limit. Our approach starts by assuming the generation of an N00N state via the first nonlinear beam splitter, leading to the following transformation of the bipartite field modes.

\begin{align} \nonumber
|N\rangle_a\otimes|\text{0}\rangle_b \equiv |\text{N,0}\rangle &\to \frac{1}{\sqrt{2}}\left(|\text{N,0}\rangle + |\text{0,N}\rangle\right), \\
|0\rangle_a\otimes|N\rangle_b \equiv|\text{0,N}\rangle &\to \frac{1}{\sqrt{2}}\left(|\text{0,N}\rangle - |\text{N,0}\rangle\right).
\label{eq24}
\end{align}

The quantum state just before the mode-sample dispersive interaction becomes---see Eq.~\eqref{eq_initial_state}:
\begin{eqnarray}
    \nonumber \rho_0 &{=}& \frac{1}{2}\left(|\text{N,0}\rangle\langle\text{N,0}|{+} |\text{N,0}\rangle\langle\text{0,N}|{+} |\text{0,N}\rangle\langle\text{N,0}|{+}|\text{0,N}\rangle\langle\text{0,N}| \right)\\
    &\otimes&\rho_\mathrm{qubits}(T).
\end{eqnarray}
During the mode-sample dispersive interaction, the system evolves according to the following.
\begin{eqnarray} \nonumber
\rho(t) &=& \frac{1}{2}\Bigg(
|N,0\rangle\langle N,0|{+}e^{-iN\chi t \hat{M}_\mathrm{qubits}} |N,0\rangle\langle 0,N|  \\
\nonumber &{+}&|0,N\rangle\langle 0,N|{+}e^{iN\chi t \hat{M}_\mathrm{qubits}} |0,N\rangle\langle N,0|\Bigg) {\otimes}\rho_\mathrm{qubits}(T).\\ \label{eq_differ_phase}
\end{eqnarray}

\begin{figure}[t]
\centering\includegraphics[width=\linewidth]{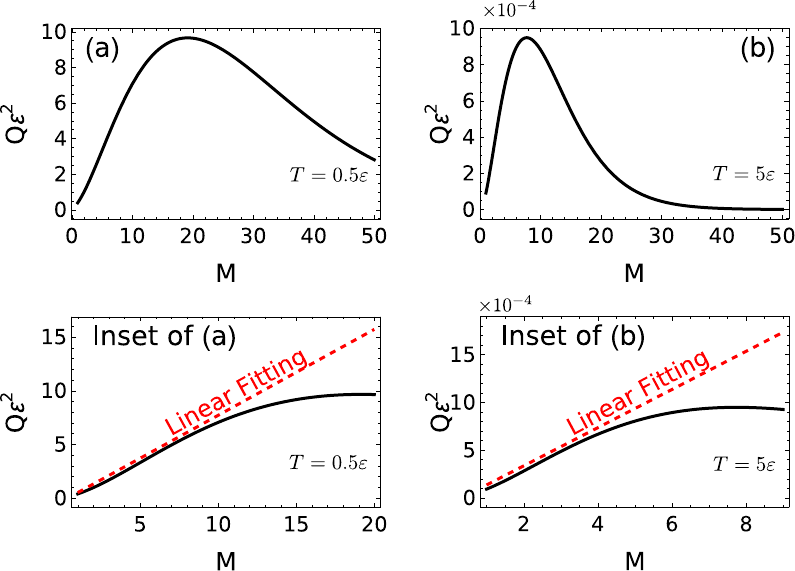}
    \caption{Quantum Fisher information (QFI) as a function of $M$ for different temperatures $T$. There is no improvement in sensing performance as the number of qubits $M$ increases, in agreement with the analytical results in Eqs.~\eqref{eq31} and~\eqref{eq_QFI_taylor}. We used $2N_\mathrm{eff}{=}1.$}
    \label{fig_qFIvnfit}
\end{figure}

By tracing out the modes $b$ and the degrees of freedom of the two-level atoms, that is, $\mathrm{Tr}_{\backslash a}$, we obtain the final state for the mode $a$:
\begin{align}
\rho_{a}(t)=p_{0}(T)|0\rangle_{a}\langle0|+p_{N}(T)|N\rangle_{a}\langle N|,
\label{eq28}
\end{align}
where
\begin{eqnarray}
&& \nonumber p_{0}(T){=}\langle \cos^{2}\left(\frac{N\chi t\hat{M}_{\text{qubits}}}{2}\right)\rangle\\
\nonumber &=& \frac{1}{Z(T)}\mathrm{Tr}\left[\cos^{2}\left(\frac{N\chi t\hat{M}_{\text{qubits}}}{2}\right)\exp\left(-\frac{\varepsilon}{T}\hat{M}_\mathrm{qubits}\right) \right],\\
\label{eq29}
\end{eqnarray}
and $p_{N}(T) = 1 - p_{0}(T)$---see the Appendix \ref{app:BS} for its analytic derivation. Given the diagonal form of Eq.~\eqref{eq29}, it is evident that the CFI, when computed on the measurement basis $\{|0\rangle\langle 0|,|N\rangle\langle N|\}$, matches the QFI. The latter can now be directly evaluated using:
\begin{equation}
Q(T) = \frac{1}{p_{0}(T)[1-p_{0}(T)]}\left[\frac{\partial p_{0}(T)}{\partial T}\right]^{2},
\label{eq30}
\end{equation}
or in closed analytical form, see Appendix \ref{app:QFI} for details:
\begin{equation}
Q(T) = \frac{M^{2}\alpha^{2M}g^{2}\cos^{2}(M\theta+\Phi)}{\left[  1-\alpha^{2M}\cos^{2}(M\theta)\right]}.
\label{eq31}
\end{equation}
In the above, the quantities $\alpha(T,N_\mathrm{eff}):=\alpha, g(T,N_\mathrm{eff}):=g, \theta(T,N_\mathrm{eff}):=\theta$ and $\Phi(T,N_\mathrm{eff}):=\Phi$ all depend on the temperature $T$ and the effective number of field excitations. 
\begin{equation}
    N_\mathrm{eff}=\frac{\varepsilon N\chi t}{2}. \label{eq_Neff}
\end{equation}
Explicitly, these quantities are defined as---see Appendix \ref{app:QFI} for details:
\begin{eqnarray}
     \theta&=&\arcsin\left(\frac{e^{-\varepsilon/T}\sin(2N_\mathrm{eff})}{\sqrt{1{+}2e^{-\varepsilon/T}\cos(2N_\mathrm{eff}){+}e^{-2\varepsilon/T}}}\right),\\     
 0<\alpha&=&\frac{\sqrt{1+2e^{-\varepsilon/T}\cos(2N_\mathrm{eff})+e^{-2\varepsilon/T}}}{1+e^{-\varepsilon/T}}\leq 1,\\
    g&=&\sqrt{C^{2} + D^{2}},\\
    \quad C &=& \frac{1}{\alpha} \frac{\partial\alpha}{\partial T},\\
    \quad D &=& -\frac{\partial\theta}{\partial T}, \\
    \Phi &=& \arctan\left(\frac{-D}{C}\right).
\end{eqnarray}
Eq.~\eqref{eq31} is the central theoretical result of our work. It directly reveals several key quantum sensing features:

Firstly, Eq.~\eqref{eq31} appears to suggest an $M^2$ dependence. However, since $0 < \alpha \leq 1$, a straightforward Taylor expansion reveals that the QFI follows a polynomial form:  
\begin{equation}
    Q(T) \sim c_1(T) M + c_2(T) M^2 + O(M^3),\label{eq_QFI_taylor}
\end{equation}
where $c_1(T)$ and $c_2(T)$ are temperature-dependent coefficients. This confirms that while an $M^2$ term exists in Eq.~\eqref{eq31}, the leading-order behavior always includes a linear $M$ term. To analyze this scaling, we fix $\alpha$ for a specific parameter choice, $N_\mathrm{eff} = 1/2$, and examine how $Q(T)$ depends on $M$. In Fig.~\ref{fig_qFIvnfit}, we plot the QFI as a function of $M$ for different temperatures $T$. The plot clearly shows that $Q(T)$ initially increases linearly with $M$. However, due to the exponential term $\alpha^{2M} $ (with $\alpha \leq 1$) , an inflection point emerges, beyond which $Q(T)$ reaches a maximum and then decreases monotonically toward zero. In conclusion, the QFI scales, at best, linearly with the number of independent two-level atoms $M$.

\begin{figure}[b]
    \centering
    \includegraphics[width=\linewidth]{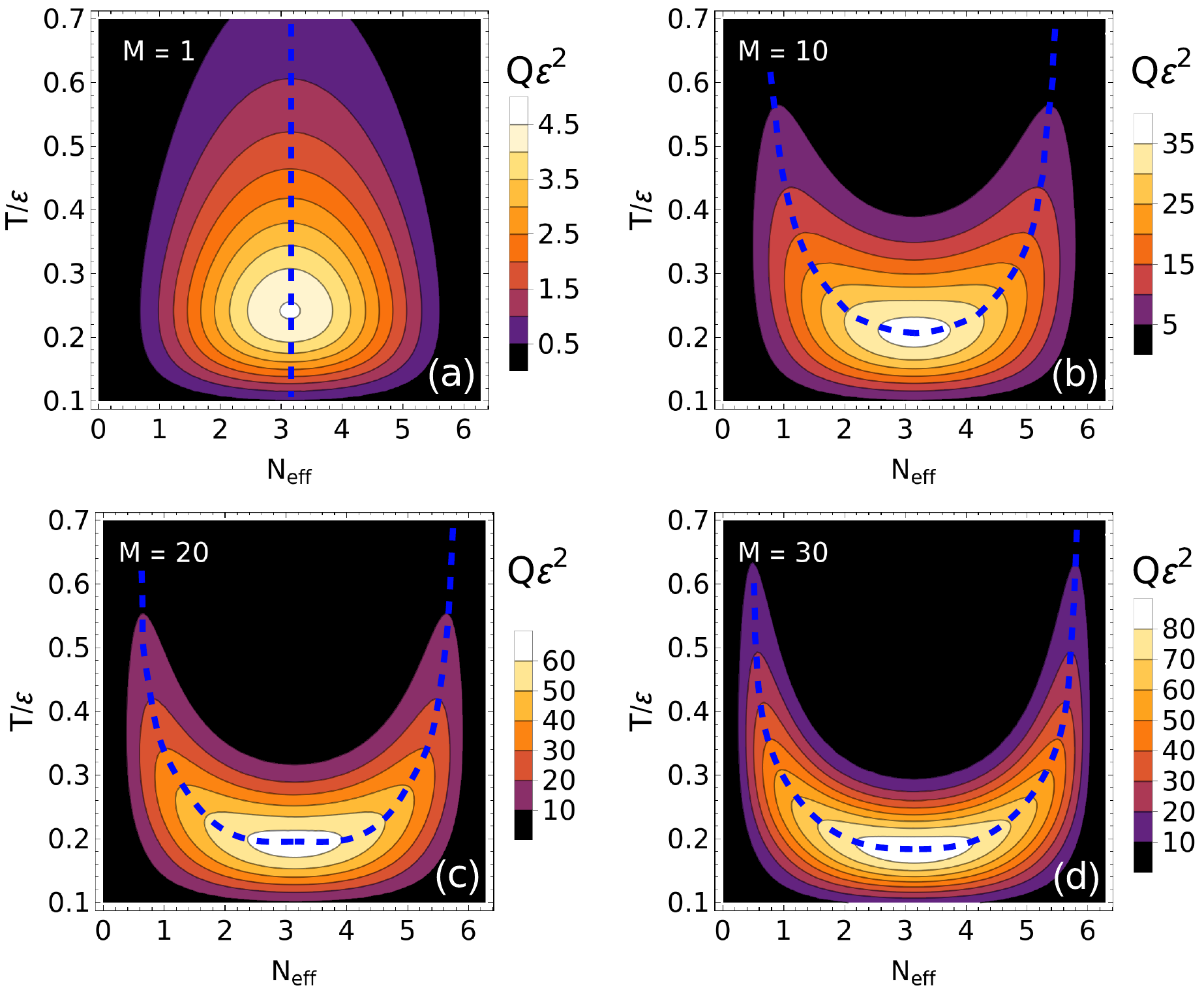}
    \caption{QFI as functions of temperature $T$ and the effective number of field excitations $N_\mathrm{eff}{=}\varepsilon N \chi t / 2$ for different numbers of independent qubits $M$. Dashed blue lines are the values of $N_\mathrm{eff}$ where the QFI reaches its maximum for a given $T$. No quantum sensing advantage is observed by increasing $N$. Other values are: $\chi{=}t{=}\varepsilon{=}1$.}
    \label{fig_QFI_TN}
\end{figure}

Secondly, in calculating the QFI in Eq.~\eqref{eq31}, we differ fundamentally from the scenario where the phase angle  in Eq.~\eqref{eq_differ_phase} is assumed to be directly proportional to the average excitation number of the thermal sample, i.e., $\phi = N\chi t \langle \hat{M}_\mathrm{qubits} \rangle$. Specifically, for $N = M$ and assuming $\phi = N\chi \langle \hat {M}_{qubits} \rangle t$, error propagation yields $Q(T) \propto N^2$ \cite{stace2010quantum}, achieving the Heisenberg limit. However, the fact that 
\begin{equation}
    \langle e^{iN\chi t \hat{M}_\mathrm{qubits}}\rangle \neq e^{iN\chi t \langle \hat{M}_\mathrm{qubits} \rangle},
\end{equation}
leads to the QFI in Eq.~\eqref{eq31}, where the thermometry limits can, at best, scale proportionally to $N$. To support this claim, Fig.~\ref{fig_QFI_TN} shows the QFI as functions of temperature $T$ and the effective number of field excitations $N_\mathrm{eff}$ for different numbers of independent two-level atoms $M$. As the figure shows, the NOON state affects the QFI solely by modulating the interaction time through the effective field excitation parameter $N_\mathrm{eff} = \varepsilon N \chi t/2$. This follows directly from Eq.~\eqref{eq31}, where $N_\mathrm{eff}$ appears only as the argument of sinusoidal functions. Importantly, the NOON state is not necessary to enhance the QFI with increasing excitation number $N$. Any apparent increase can instead be achieved by appropriately adjusting the interaction coupling strength $\chi$ or the interaction time $t$ for a given $N$. In addition, in Fig.~\ref{fig_QFI_TN}, the dashed blue lines indicate the values of $N_\mathrm{eff}$ where the QFI reaches its maximum for a given $T$. Within the range $0 \leq N_\mathrm{eff} \leq 2\pi$, there are two such points for each $T$. Beyond this range, the QFI shows a periodic pattern.

\begin{figure}[t]
    \centering
    \includegraphics[width=\linewidth]{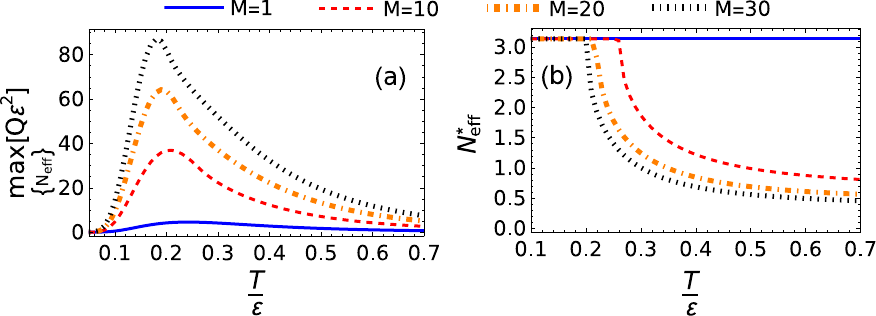}
    \caption{(a) QFI maximized over $N_\mathrm{eff}$ as a function of $T$ for several values of number of qubits $M$. (b) Values of $N^*_\mathrm{eff}$ that maximize the QFI for a given $T$.}
    \label{fig_maxQFI_T}
\end{figure}

Lastly, in Fig.~\ref{fig_maxQFI_T}(a), we show the QFI maximized over $N_\mathrm{eff}$ as a function of $T$ for several values of number of qubits $M$. Each curve corresponds to the blue dashed line in Fig.~\ref{fig_QFI_TN}. As seen from Fig.~\ref{fig_maxQFI_T}(a), no clear quantum sensing advantage is shown by increasing $N$. In fact, in Fig.~\ref{fig_maxQFI_T}(b), we plot the values of $N^*_\mathrm{eff}$ that maximize the QFI for each $T$. The figure shows that the optimal $N_\mathrm{eff}$ even decreases with $T$. Note that this optimization can be controlled either by adjusting the interaction time $t$ or by modifying the N00N state excitation.

\section{The experimental setup}\label{sec_experimental}

This section presents the experimental implementation corresponding to our theoretical thermometry model. We begin by introducing the quantum circuit used to simulate the theoretical predictions for arbitrary values of relevant parameters. We then describe the actual experiment, which was performed using $M$ two-level atoms and a N00N state with a single excitation ($N=1$), since, as can be seen in Eq. \eqref{eq_Neff}, $N$ just rescales the interaction time.

To simulate the MZI shown in Fig.~\ref{fig_sketch_MZ}, we use Qiskit~\cite{qiskit} to construct quantum circuits that represent the interferometer's operation. These circuits are then executed on the IBM\_brisbane quantum processor via the IBM Quantum Platform~\cite{ibmq}. As explained below, the simulation models a MZI containing $M$ independent atoms in one arm interacting with a maximally entangled bipartite field state. 

\subsection{Qubit State Preparation in Gibbs Form}

To prepare each qubit in a Gibbs state at temperature $T$, given by  
\begin{equation}
\rho_\mathrm{qubits}(T) = Z(T)^{-1} e^{ -\varepsilon/T \, \hat{\sigma}_+^{(1)} \hat{\sigma}_-^{(1)} },
\end{equation}
we require two qubits per site and used $M=1$ in Eq.~\eqref{eq_Hqubits}. One qubit serves as an ancilla, while the other represents the qubit whose thermal state we wish to prepare at an unknown temperature.

\begin{figure}[t]
\centering
\includegraphics[width=\linewidth]{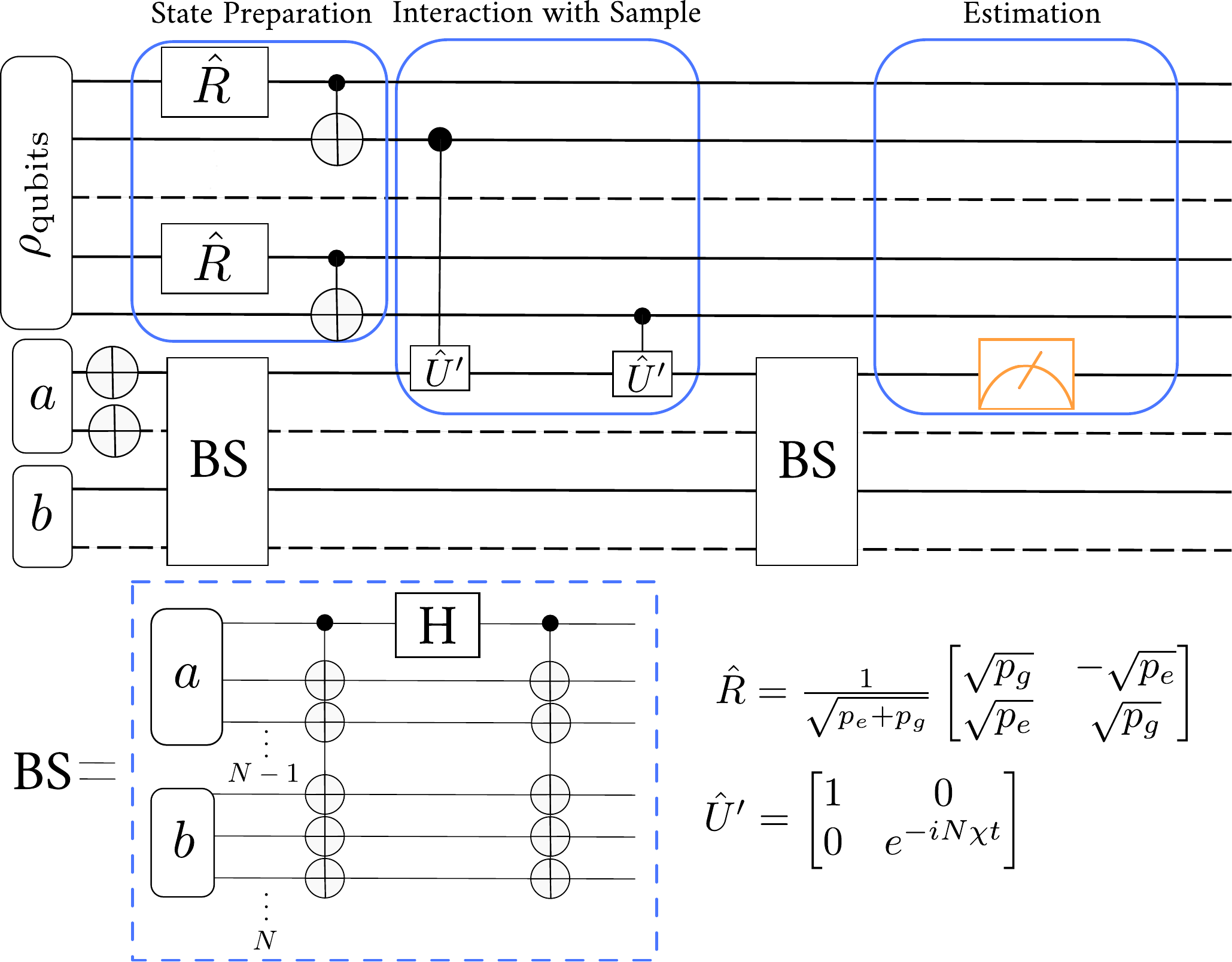}
\caption{Circuit scheme for parameter estimation. Registers $\bm{a}$ and $\bm{b}$ represent the two interferometer arms, while $\mathbf{\rho}_\mathrm{qubits}$ denotes the register where a mixed thermal state of $M$ independent qubits is prepared. In the ``State Preparation" (blue box), the qubits are initialized with the aid of ancillary qubits. A nonlinear beam splitter (dashed blue box) is constructed using multiple-controlled-X and Hadamard gates to generate the required N00N state, $|N00N\rangle = (\ket{0,N} + \ket{N,0})/\sqrt{2}$. In the ``Interaction with Sample" (blue box), a 
 sequence of controlled unitaries simulates the dispersive coupling between the qubits and field mode $\bm{a}$---see Eq.~\eqref{eq_H_interaction}. Finally, in the ``Estimation" (blue box), measurements on the computational $\hat{\sigma}_z$ basis yield statistics from which the temperature-dependent probability distribution is extracted.}
\label{fig_quantum_circuit}
\end{figure}

To achieve this, we apply the unitary single-qubit rotation gate
\begin{eqnarray}
\hat{R}(\Omega)&=&\begin{bmatrix}
        \cos\Omega &-\sin\Omega\\
        \sin\Omega &\cos\Omega
    \end{bmatrix}\\
    &=&\frac{1}{\sqrt{p_e(\Omega){+}p_g(\Omega)}}\begin{bmatrix}
        \sqrt{p_g(\Omega)} &-\sqrt{p_e(\Omega)}\\
        \sqrt{p_e(\Omega)} &\sqrt{p_g(\Omega)}
    \end{bmatrix},
\end{eqnarray}
where the freely tunable rotation angle $\Omega$ parametrizes the probabilities $p_e(\Omega)$ and $p_g(\Omega)$ via $\cos\Omega{=}\sqrt{p_g(\Omega)}$ and $\sin\Omega{=}\sqrt{p_e(\Omega)}$. By construction, these definitions ensure that $\{p_e(\Omega), p_g(\Omega)\}{\geq}0$ and that the sum $p_e(\Omega){+}p_g(\Omega){=}1$. For brevity, we will omit the explicit dependence on $\Omega$ from this point forward. By applying $\hat{R}$ to the ancilla qubit, we get:
\begin{equation}
    \hat{R} |g\rangle =  \sqrt{p_g} |g\rangle + \sqrt{p_e} |e\rangle.
\end{equation}
Hence, it is straightforward to obtain:
\begin{equation}
    \rho_{1,1'} = \text{cNOT}_{1,1'}(\hat{R}_1\otimes I_{1'}) |g,g\rangle_{1,1'}\langle g,g|(\hat{R}^\dagger_1\otimes I_{1'})\text{cNOT}_{1,1'}^\dagger,
\end{equation}
where $\text{cNOT}_{1,1'}$ is the cNOT two-qubit gate. Therefore, by tracing out the ancilla $1'$ from the above joint state $\hat{\rho}_{1,1'}$, we obtain the reduced state of the qubit:
\begin{equation}
    \rho_1 \equiv \rho_\mathrm{qubit} = p_e|e\rangle_1\langle e| + p_g |g\rangle_1\langle g|.
\end{equation}
See ``State Preparation" in Fig.~\ref{fig_quantum_circuit} for the corresponding quantum circuit. This procedure is applied independently to each qubit in the system. To prepare $M$ independent qubits in a Gibbs state, a total of $2M$ physical qubits are required. Note that the rotation angle $\Omega$ determines the thermal occupation probabilities through the mapping:
\begin{eqnarray}
p_g &=& \frac{1}{1 + e^{-\varepsilon/T}},\\
p_e &=& \frac{e^{-\varepsilon/T}}{1 + e^{-\varepsilon/T}}.
\end{eqnarray}
Thus, selecting an appropriate value of $\Omega$ effectively encodes the corresponding Gibbs distribution, allowing the rotation to simulate thermalization at temperature $T$.

\subsection{NOON State preparation}

To experimentally create the N00N state that will interact with the sample of $M$ qubits, we simulate a nonlinear beam splitter by employing a combination of multiple-controlled-X (MCX) gates and a Hadamard (H) gate. For consistency with previous sections, we adopt the computational basis $\{|0\rangle,|1\rangle \}$ to represent the field excitations associated with the N00N states. This choice allows us to distinguish the system having $N$ excitations from the sample of $M$ qubits (in basis $\{|e\rangle,|g\rangle\}$) at an unknown temperature $T$.

To illustrate the procedure, consider the specific case of $N = 3$ excitations, where the desired N00N state is:
\begin{equation}
    |3,0\rangle \longrightarrow \frac{1}{\sqrt{2}} \left( |0,3\rangle - |3,0\rangle \right).
\end{equation}
We define the states $|3\rangle \equiv |1,1,1\rangle$ and $|0\rangle \equiv |0,0,0\rangle$. Let us initialize the system in the state: $|\Psi\rangle = |1\rangle \otimes |1,1,0,0,0\rangle$, where the leftmost qubit $|1\rangle$ acts as the control qubit, and the remaining five qubits $|1,1,0,0,0\rangle$ are the target qubits. We first apply a MCX gate to $|\Psi\rangle$, flipping the qubits as $\text{MCX}|\Psi\rangle = |1\rangle \otimes |0,0,1,1,1\rangle$. We now apply a Hadamard gate $\text{H}$ to the control qubit---recall that $\text{H}|0\rangle = \frac{1}{\sqrt{2}}(|0\rangle + |1\rangle)$ and $\text{H}|1\rangle = \frac{1}{\sqrt{2}}(|0\rangle - |1\rangle)$. After applying $\text{H}$, the state becomes: $\text{H}(\text{MCX}|\Psi\rangle) = \frac{1}{\sqrt{2}} \left( |0\rangle - |1\rangle \right) \otimes |0,0,1,1,1\rangle$. Finally, we apply another MCX gate, which flips each of the target qubits conditioned on the control qubit being in state $|1\rangle$. This operation transforms the state as follows:
\begin{equation}
 \frac{1}{\sqrt{2}} \left( |0\rangle \otimes |0,0,1,1,1\rangle - |1\rangle \otimes |1,1,0,0,0\rangle \right).    
\end{equation}
Identifying $|0\rangle \otimes |0,0,1,1,1\rangle \equiv |0,3\rangle$ and $|1\rangle \otimes |1,1,0,0,0\rangle \equiv |3,0\rangle$, we arrive at the entangled N00N state:
\begin{equation}
|3,0\rangle \rightarrow \frac{1}{\sqrt{2}} \left( |0,3\rangle - |3,0\rangle \right).    
\end{equation}
This procedure effectively implements the needed nonlinear beam splitter operation, see BS diagram in Fig.~\ref{fig_quantum_circuit}. Similarly, we can initialize the system in the state: $|\Psi\rangle = |0\rangle \otimes |0,0,1,1,1\rangle\equiv|0,3\rangle$ and then create the state $\frac{1}{\sqrt{2}} \left( |0,3\rangle + |3,0\rangle \right)$---or alternatively, we can introduce an additional relative phase (e.g., using a controlled-Z gate).

\subsection{Quantum Temporal Evolution}

To simulate the system dynamics on a digital quantum simulator, we employ quantum controlled-unitary (cU) gates. In this setup, the control is provided by the state of the qubit (a two-level atom from the sample), while the target is the mode $a$ in the MZI. This approach becomes evident by examining the time-evolution operator $\hat{U}(t) = e^{-it\hat{H}}$, where the Hamiltonian $\hat{H}$ is defined in Eq.~\eqref{eq_H_interaction}. Indeed, this global unitary $\hat{U}(t)$ can be factorized as a product of local unitaries:
\begin{equation}
\label{eq:41}
\hat{U}(t) = \prod_{k=1}^M \opi{u}_k(t),    
\end{equation}
where each $\opi{u}_k(t)$ acts on the $k$th qubit and the field mode $a$, and is simply given by
\begin{equation}
\opi{u}_k(t) = e^{-it\chi  \opi{a}^\dagger \opi{a}\hat{\sigma}_+^{(k)} \hat{\sigma}_-^{(k)}}.
\label{eq:42}
\end{equation}
The above expression shows that $\opi{u}_k(t)$ corresponds to the time evolution operator for the interaction between the $k$th qubit and the mode $a$, and has the structure of a controlled operation: the evolution of the mode $a$ is conditioned on the $k$th qubit being in its excited state.
  
For the $k-$th interaction, the density matrix of our mixed system allows the following possibilities for the system states: $\ket{0}_a\ket{g}_k$, $\ket{0}_a\ket{e}_k$, $\ket{N}_a\ket{g}_k$ and $\ket{N}_a\ket{e}_k$. It is straightforward to notice that the action of $\opi{u}_k(t)$ given by Eq. \eqref{eq:42} on these states is:
\begin{equation}
\label{eq:43}
\begin{aligned}
&\opi{u}_k(t)\ket{0}_a\ket{g}_k=\ket{0}_a\ket{g}_k;\quad \opi{u}_k(t)\ket{0}_a\ket{e}_k=\ket{0}_a\ket{e}_k\\
&\opi{u}_k(t)\ket{N}_a\ket{g}_k=\ket{N}_a\ket{g}_k;\quad \opi{u}_k(t)\ket{N}_a\ket{g}_k=e^{-i\chi tN}\ket{N}_a\ket{g}_k. 
\end{aligned}
\end{equation}

Now consider the following operator acting on the light mode a:
\begin{equation}
\hat{U}'\ket{0}_a=\ket{0}_a;\quad \hat{U}'\ket{N}_a=e^{-i\chi Nt}\ket{N}_a.
\end{equation}
So the matrix representation of $\hat{U}'$ is:
\begin{equation}
\label{eq:45}
\hat{U}'=
\begin{bmatrix}
1&0\\
0&e^{-iN\chi t}
\end{bmatrix},
\end{equation}
and its controlled form (with $k-$th qubit serving as control) $c\hat{U}'_k$ is:
\begin{equation}
c\hat{U}'_k=\opi{I}_a\otimes|g\rangle\langle g|_k  +\hat{U}' \otimes |e\rangle\langle e|_k  
\end{equation}
Then, the action of $c\hat{U}'_k$ on the possible system states is given by:
\begin{equation}
\label{eq:47}
\begin{aligned}
&c\hat{U}'_k\ket{0}_a\ket{g}_k=\ket{0}_a\ket{g}\braket{g|g}_k+\hat{U}'\ket{0}_a\ket{e}\braket{e|g}_k=\ket{0}_a\ket{g}_k\\
&c\hat{U}'_k\ket{0}_a\ket{e}_k=\ket{0}_a\ket{g}\braket{g|e}_k+\hat{U}'\ket{0}_a\ket{e}\braket{e|e}_k=\ket{0}_a\ket{e}_k\\
&c\hat{U}'_k\ket{N}_a\ket{g}_k=\ket{N}_a\ket{g}\braket{g|g}_k+\hat{U}'\ket{N}_a\ket{e}\braket{e|g}_k=\ket{N}_a\ket{g}_k\\
&c\hat{U}'_k\ket{N}_a\ket{e}_k{=}\ket{N}_a\ket{g}\braket{g|e}_k{+}\hat{U}'\ket{N}_a\ket{e}\braket{e|e}_k{=}e^{-i\chi t N}\ket{N}_a\ket{e}_k.\\
\end{aligned}
\end{equation}
As can be seen by comparison of Eqs. \eqref{eq:47} and \eqref{eq:43}, the action of $c\hat{U}_k$ is equivalent to the action of $\opi{u}_k(t)$. Therefore, $\hat{U}(t)$ in Eq. \eqref{eq:41} can be simulated through the sequence of controlled operations depicted in the "interaction with the sample" box in Fig. \ref{fig_quantum_circuit}, because since $c\hat{U}'_k=\opi{u}_k(t)$, we can establish:
\begin{equation}
\label{eq:48}
\hat{U}(t)=\prod_{k=1}^Mc\hat{U}'_k 
\end{equation}

\subsection{Implementation}

We now have all the necessary ingredients to quantum simulate a thermometry scheme based on nonlinear MZI. As previously discussed, increasing the photon number $N$ does not provide any quantum enhancement in sensitivity---it merely rescales the interaction time or coupling strength. Therefore, we focus on the cases $N = 1, 2$ and $3$ to explore thermometry performance, while varying the number of qubits $M$, since increasing $M$ improves sensitivity with quantum standard limit scaling.

For each experiment, we perform $5 \times 10^3$ independent measurement runs in the computational $\hat{\sigma}_z$ basis at a fixed temperature $T$ (encoded in the rotation angle $\Omega$). The measurement outcomes are used to construct the probability distributions $p_0$ and $p_1$, from which we directly compute a single value of the CFI. As previously discussed, in this case, the CFI and the QFI are equal. This entire procedure is repeated  across 30 independent experiments, allowing us to calculate the mean and standard deviation of the Fisher information and its fluctuations.

To demonstrate the flexibility in preparing the qubit states, we also consider qubits initialized at effective negative temperatures, namely with inverted population, where the excited state is more populated than the ground state. This is physically allowed because the qubit spectrum is discrete, and all qubits are identical and non-interacting. Accordingly, we explore the thermometry capabilities for both positive and effective negative temperatures.

The complete batch of circuits is sent for execution on the IBM\_brisbane QPU, with an optimization level of 3 and 5000 shots per circuit. All quantum circuits are designed with the quantum probe that maximizes the QFI.

\section{Experimental Results}

The quantum circuit in Fig.~\ref{fig_quantum_circuit} enables the preparation of arbitrary qubit mixed states. This includes states with population inversion, where the excited state is more populated than the ground state---such states correspond to effective negative temperatures~\cite{ramsey1956thermodynamics}. To readily observe this, let us compute the average excitation, for one of the $M$ identical and individual two-level atoms described by the Hamiltonian in Eq.~\eqref{eq_Hqubits}, $\langle{\hat{\sigma}^{+}\hat{\sigma}^{-}}\rangle=\mathrm{Tr}[{\hat{\sigma}^{+}\hat{\sigma}^{-}}(p_g|g\rangle\langle g| + p_e|e\rangle\langle e|)]$ as:
\begin{equation}
\langle{\hat{\sigma}^{+}\hat{\sigma}^{-}}\rangle=p_{e}=\frac{1}{e^{\varepsilon/T}+1}.
\label{eq17}
\end{equation}
The above expression follows the Fermi-Dirac distribution---recall that $\varepsilon > 0$. Note that $0\leq p_e \leq 1$, with $p_e = \frac{1}{2}$ in the limit $T/\varepsilon \to \pm\infty$. Indeed, for infinitely large positive temperatures $T/\varepsilon \to +\infty$, the excited-state population is $p_e=\frac{1}{2}$. As the temperature decreases toward zero from above $T/\varepsilon \to 0^+$, $p_e \to 0$, meaning full ground-state occupation. In contrast, as $T/\varepsilon \to 0^-$, the population inverts and $p_e \to 1$, while for $T/\varepsilon \to -\infty$, $p_e =\frac{1}{2}$ again. Therefore, the Fermi-Dirac distribution ensures that $p_e < \frac{1}{2}$ for $T/\varepsilon > 0$, $p_e > \frac{1}{2}$ for $T/\varepsilon \to 0$, and $p_e = \frac{1}{2}$ in the limit $T/\varepsilon \to \pm\infty$. Effective negative temperatures are thus understood as states exhibiting population inversion.

\begin{figure}[t]
\centering
 \includegraphics[width=\linewidth]{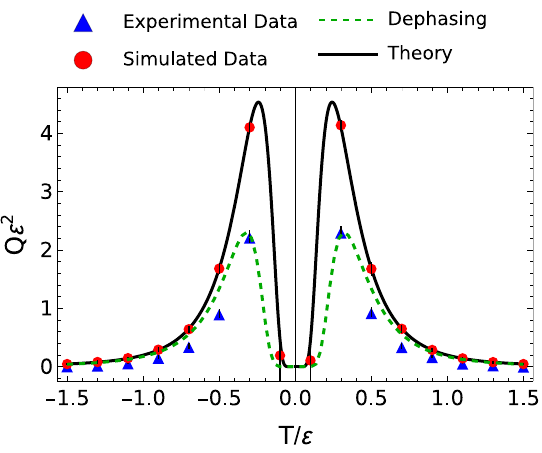}
 \caption{Comparison of the QFI as a function of temperature $T$ for the case $N = 1$, $M = 1$. Red circles represent the simulated QFI. Blue triangles correspond to experimental data. Black solid line shows the ideal theoretical prediction. Green dashed line represents the theoretical prediction under depolarizing noise.}
\label{fig_QFI_experiment}
\end{figure}

In Fig.~\ref{fig_QFI_experiment}, we compare the simulated, experimental, and theoretical QFI as a function of temperature for the case $N=1$ and $M=1$. The figure shows that the QFI is symmetric with respect to both positive and effective negative temperatures. This symmetry arises because, for a two-level system, the heat capacity is symmetric under the transformation $T\to -T$. To understand the behavior of the theoretical QFI in Fig.~\ref{fig_QFI_experiment}, we analyze two limiting cases. First, in the low-temperature limit $|T/\varepsilon| \to 0$, the qubit remains in its ground state. In this regime, the probe's state no longer depends on the unknown temperature $T$, and the QFI therefore vanishes. Second, in the high-temperature limit $|T/\varepsilon| \gg 1$, the probe becomes fully thermalized: the excited and ground states are equally populated ($p_e = p_g = 1/2$), corresponding to $\rho_\mathrm{qubit}=I/2$. In this case, the probe also becomes insensitive to small changes in $T$, and again the QFI drops to zero.
Fig.~\ref{fig_QFI_experiment} demonstrates that, for most data points, the simulated QFI closely matches the theoretical QFI curve. In the region where $|T/\varepsilon| < 1/2$, the average simulated QFI aligns well with the theoretical prediction. However, as the temperature approaches zero $T \to 0$, the statistical uncertainty in the simulation results increases noticeably.

Finally, note from Fig.~\ref{fig_QFI_experiment}, that the overall trend of the experimental data follows the theoretical curve, while the magnitude of the experimental QFI is consistently lower. This reduction can be attributed to the presence of noise and decoherence effects that are unavoidable in real experimental settings. Among these, a key contributor is the action of a depolarizing channel. To account for this effect, we model the probe as follows:
\begin{equation}
    \rho(t) \to p_\gamma \rho(t) + (1 - p_\gamma)\frac{I}{2}.
\end{equation}
The expression above represents a statistical mixture, where the quantum state $\rho$ is prepared with probability $0\leq p_\gamma \leq 1$ and mixed with the identity with probability $1 - p_\gamma$. As shown in Fig.~\ref{fig_QFI_experiment}, we find that introducing a modest depolarizing probability of $p_\gamma = 0.95$ is sufficient to reduce the theoretical QFI to a level that closely matches the experimental data.

\begin{figure}[t]
\centering
 \includegraphics[width=\linewidth]{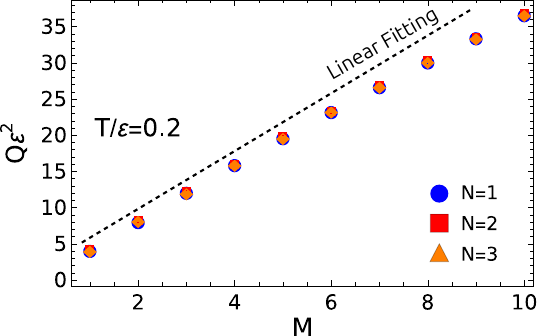}
 \caption{Digitally simulated QFI as a function of $M$ at $T=0.2\varepsilon$ for several values of $N$.}
\label{fig_QFI_NM}
\end{figure}

To further support our claim that, in general, N00N states do not provide a quantum advantage for thermometry under dispersive interaction, we show in Fig.~\ref{fig_QFI_NM} the digitally quantum simulated QFI as a function of $M$ at $T=0.2\varepsilon$ for the values of $N = 1, 2, 3$, and maximized independently for each given pair $(N, M)$. As shown in the figure, the QFI curve overlaps across different $N$ values, indicating that: (i) $N$ merely rescales the effective parameters $t$ and $\chi$ as described in Eq.~\eqref{eq_Neff}, and (ii) our digital quantum simulation performs well, especially in accurately implementing the nonlinear beam splitter model used to generate N00N states with arbitrary $N$. Furthermore, Fig.~\ref{fig_QFI_NM} clearly demonstrates that the QFI scales---at best---according to the standard quantum limit, confirming the absence of Heisenberg scaling and thus the lack of  quantum enhancement either from using N00N sates or increasing the number $M$ of two-level atoms in this setting.

\section{Conclusion}

In this work, we have two main contributions: First, unlike previous claims based on strict assumptions, we theoretically demonstrate---without making any approximations or assumptions about the quantum state---that the unknown temperature encoded in the correlated atom-light state can, at best, be estimated with a precision scaling at the standard quantum limit. Second, we propose a non-linear Mach-Zehnder interferometric setup suitable for implementation on digital quantum simulators. We experimentally show that the standard quantum limit is indeed achieved in the single-photon excitation subspace and verify, through digital simulations, that this precision scaling persists for larger photon numbers and atomic excitations. We show that our approach also applies to atomic ensembles prepared at both positive and effective negative temperatures, highlighting the flexibility and generality of our method. The proposed implementation, together with the theoretical uncertainty bounds, is highly relevant for quantum technologies in the Noisy Intermediate-Scale Quantum era, where thermal fluctuations are ubiquitous.

\begin{appendix}

\section{Nonlinear Mach-Zehnder Interferometer and Interaction with M Thermal Two-Level Atoms}\label{app:BS}

Assuming there are $M$ non-interacting two-level atoms in the sample, we can write the total thermal state as the product of thermal states for two-level systems:
\begin{equation}
\opi{\rho}_{T} = \sum_{n_{1},n_{2},\ldots,n_{M}=0,1} p_{n_{1}} p_{n_{2}} \cdots p_{n_{M}} |n_{1},n_{2},\ldots,n_{M}\rangle\langle n_{1},n_{2},\ldots,n_{M}|,
\end{equation}
where $p_{n_{k}}$ is the probability of each qubit $k$ being in the state $n_{k}=0$ or $1$:
\begin{equation}
p_{n_{k}=1} = \frac{e^{-\varepsilon/T}}{1+e^{-\varepsilon/T}}, \quad p_{n_{k}=0} = \frac{1}{1+e^{-\varepsilon/T}},
\end{equation}
with $p_{n_{k}=1} + p_{n_{k}=0} = 1$ and $|n_{1},n_{2},\ldots,n_{M}\rangle$ representing the tensor product in the entire Hilbert space of $M$ qubits. Here, $\varepsilon$ is the energy spacing between the excited state and excited state of an individual two-level systems. Since the interaction between the thermal state and the state in the interferometer is dispersive, we will keep it until the end as a tensor product with the beam splitter (BS) states.

\subsection{Nonlinear Beam Splitter Evolution}

Now we assume that the nonlinear BS produces the following evolution
\begin{equation}
|N,0\rangle \to \frac{1}{\sqrt{2}}\left(|N,0\rangle + |0,N\rangle\right), |0,N\rangle \to \frac{1}{\sqrt{2}}\left(|0,N\rangle - |N,0\rangle\right).
\end{equation}
Therefore, after the first BS the state is $\opi{\rho}_{\text{NOON}}\otimes\opi{\rho}_{T}  $, with $\opi{\rho}_{\text{NOON}} = |\text{NOON}\rangle\langle\text{NOON}|$ and
\begin{equation}
|\text{NOON}\rangle = \frac{1}{\sqrt{2}}\left(|N,0\rangle_{ab} + |0,N\rangle_{ab}\right).
\end{equation}

The interaction Hamiltonian is ($\hbar = 1$):
\begin{equation}
\opi{H} =\varepsilon \chi \opi{a}^{\dagger}\opi{a} \sum_{k=1}^{M} \opi{\sigma}_{+}^{(k)}\opi{\sigma}_{-}^{(k)} \equiv \varepsilon\chi \opi{a}^{\dagger}\opi{a} \opi{M}_{qubits}.
\end{equation}
Note that for the NOON state we have $\opi{a}^{\dagger}\opi{a} |N,0\rangle_{ab} = N |N,0\rangle_{ab}$. Therefore, when acting on the NOON state, the operator $\opi{a}^{\dagger}\opi{a}$ effectively generates a factor $N$ in the phase acquired by the unitary evolution $\opi{U} = e^{-i\opi{H}t}$.

After time $t$, the time evolution of the total density operator $\opi{\rho}_{\text{total}} = \opi{\rho}_{\text{NOON}}\otimes\opi{\rho}_{T}  $ is given by:
\begin{equation}
\opi{\rho}_{\text{total}}(t) = e^{-i\opi{H}t} \opi{\rho}_{\text{total}} e^{i\opi{H}t}.
\end{equation}
Substituting the Hamiltonian, we find that the coherence terms $|N,0\rangle\langle 0,N|$ and $|0,N\rangle\langle N,0|$ in the NOON state gain a factor dependent on $\hat{M}_{qubits} = \sum_{k=1}^{M} \opi{\sigma}_{+}^{(k)}\opi{\sigma}_{-}^{(k)}$,
with the additional factor $N$ according to the excitation of the NOON state:
\begin{equation}
\begin{aligned}
&e^{-i\opi{H}t} |N,0\rangle_{ab}\langle 0,N| \otimes\opi{\rho}_T e^{i\opi{H}t} \\
&= e^{-i\varepsilon \opi{N}\chi t \opi{M}_\mathrm{qubits}} |N,0\rangle_{ab}\langle 0,N| \otimes \opi{\rho}_Te^{i\varepsilon \opi{N}\chi t \hat{M}_\mathrm{qubits}}\\
&=(\opi{I}_{ab}\otimes e^{-i\varepsilon N\chi t \hat{M}_\mathrm{qubits}})(|N,0\rangle_{ab}\langle 0,N| \otimes \opi{\rho}_T)(\opi{I}_{ab}\otimes e^{i\varepsilon 0\chi t \hat{M}_\mathrm{qubits}})\\
&=(\opi{I}_{ab}\otimes e^{-i\varepsilon N\chi t \hat{M}_\mathrm{qubits}}) (|N,0\rangle_{ab}\langle 0,N|\otimes \opi{\rho}_T),
\end{aligned}
\end{equation}
with $\opi{I}_{ab}$ being the identity operator in the light modes in the arms corresponding subspace.

A similar evolution occurs for $\ket{0,N}_{ab}\bra{N,0}$:
\begin{equation}
\label{eq:8}
e^{-i\opi{H}t} |0,N\rangle_{ab}\langle N,0| \otimes\opi{\rho}_T e^{i\opi{H}t}=(\opi{I}_{ab}\otimes e^{i\varepsilon N \chi t \hat{M}_\mathrm{qubits}})|0,N\rangle_{ab}\langle N,0|\opi{\rho}_T.
\end{equation}
These actions can be checked by expanding $e^{i\opi{H}t}=e^{i\varepsilon \chi t\opi{I}_b\otimes\opi{a}^{\dagger}\opi{a}\otimes \hat{M}_\mathrm{qubits}}$ as a Taylor series and taking the Kronecker products.
It is important to mention that the multiplication by $\opi{I}_{ab}\otimes e^{i\varepsilon N\chi t\hat{M}_\mathrm{qubits}}$ in Eq. \eqref{eq:8} only can be applied by the left instead of right because $[e^{i\varepsilon N\chi t \hat{M}_\mathrm{qubits}},\opi{\rho}_T]=0$, since both $\hat{M}_\mathrm{qubits}$ and $\opi{\rho}_T$ are diagonalized by the same basis.   

Also, doing the aforementioned expansions, it is possible to check that $\ket{0,N}_{ab}\bra{0,N}$ and $\ket{N,0}_{ab}\bra{N,0}$ will remain unchanged because the exponential factors at left and right sides cancel each other. From now on, the Kronecker product symbol in $(\opi{I}_{ab}\otimes e^{i\varepsilon N \chi t \hat{M}_\mathrm{qubits}})$ will be omitted for simplicity and brevity, so:
\begin{equation}
\opi{I}_{ab}\otimes e^{\pm i\varepsilon N \chi t \hat{M}_\mathrm{qubits}}\equiv e^{\pm i\varepsilon N \chi t \hat{M}_\mathrm{qubits}}
\end{equation}

With these considerations, the total evolved state becomes:
\begin{equation}
\begin{aligned}
&\opi{\rho}_{abT}(t)=\frac{1}{2}\Bigg(
|N,0\rangle_{ab}\langle N,0| + |0,N\rangle_{ab}\langle 0,N|\\
&+ e^{-i\varepsilon N\chi t \hat{M}_\mathrm{qubits}} |N,0\rangle_{ab}\langle 0,N| 
 +e^{i\varepsilon N\chi t \hat{M}_\mathrm{qubits}} |0,N\rangle_{ab}\langle N,0|\Bigg)\otimes\opi{\rho}_T.
 \end{aligned}
\end{equation}

At this point, it is important to compare this result with that of Ref.~\cite{stace2010quantum}, where the phase after the atom-thermal sample interaction is assumed to be $\phi_{B} =  \chi \langle \hat{M}_\mathrm{qubits} \rangle t$:
\begin{equation}
\opi{U}(t)|\text{NOON}\rangle=\frac{1}{\sqrt{2}}\left(|N,0\rangle_{ab}+e^{iN\phi_{B}}|0,N\rangle_{ab}\right).
\end{equation}

The difference between the results, therefore, lies in assuming that the acquired phase is proportional to $\langle \hat{M}_\mathrm{qubits} \rangle$, unlike what the direct calculation indicates, that is
\begin{equation}
\left\langle e^{iN\chi t\hat{M}_\mathrm{qubits}}\right\rangle \neq e^{iN\chi t\left\langle \hat{M}_\mathrm{qubits}\right\rangle} = e^{iN\phi_{B}}.
\end{equation}

\subsection{State After the Second BS}

After the second BS and grouping the similar terms, we find:
\begin{equation}
\begin{aligned}
&\opi{\rho}_{abT}(t)
= \frac{1}{4}\Bigg[\left(2 + e^{i\varepsilon N\chi t \hat{M}_\mathrm{qubits}} + e^{-i\varepsilon N\chi t \hat{M}_\mathrm{qubits}}\right)|0,N\rangle_{ab}\langle 0,N| +\\ 
&(e^{i \varepsilon N \chi t \hat{M}_\mathrm{qubits}}-e^{-i \varepsilon N \chi t \hat{M}_\mathrm{qubits}})|0,N\rangle_{ab}\langle N,0| \\
&+(e^{-i \varepsilon N \chi t \hat{M}_\mathrm{qubits}}-e^{i \varepsilon N \chi t \hat{M}_\mathrm{qubits}})|N,0\rangle_{ab}\langle 0,N|\\
&+ \left(2 - e^{i\varepsilon N\chi t \hat{M}_\mathrm{qubits}} - e^{-i \varepsilon N\chi t \hat{M}_\mathrm{qubits}}\right)|N,0\rangle_{ab}\langle N,0| \Bigg] \otimes \rho_{T}.
\end{aligned}
\end{equation}

Using the Euler formula and the identity for cosine, the reduced state arriving at detector $a$, while maintaining the thermal state, is given by:
\begin{eqnarray}
\opi{\rho}_{aT}(t)&=& \Big[\cos^{2}\left(\frac{\varepsilon N\chi t \hat{M}_\mathrm{qubits}}{2}\right)|0\rangle_{a}\langle 0| \\
&+& \sin^{2}\left(\frac{\varepsilon N\chi t \hat{M}_\mathrm{qubits}}{2}\right)|N\rangle_{a}\langle N|\Big] \otimes \rho_{T},
\end{eqnarray}
where we still use the compact notation:
\begin{equation}
\opi{I}_{a}\otimes \cos^{2}\left(\frac{\varepsilon N\chi t \hat{M}_\mathrm{qubits}}{2}\right) \equiv \cos^{2}\left(\frac{\varepsilon N\chi t \hat{M}_\mathrm{qubits}}{2}\right).
\end{equation}

Finally, taking the trace over the thermal state, we obtain:
\begin{eqnarray}
\opi{\rho}_{a}(t) &=& \langle \cos^{2}\left(\frac{\varepsilon N\chi t \hat{M}_\mathrm{qubits}}{2}\right) \rangle |0\rangle_{a}\langle 0| \\
&+& \langle \sin^{2}\left(\frac{\varepsilon N\chi t \hat{M}_\mathrm{qubits}}{2}\right) \rangle |N\rangle_{a}\langle N|,
\end{eqnarray}
or simply 
\begin{equation}
\opi{\rho}_{a}(t) = p_{0}|0\rangle_{a}\langle0| + p_{N}|N\rangle_{a}\langle N|,
\end{equation}
with
\begin{align} \nonumber
p_{0} & = \langle\cos^{2}\left(\frac{\varepsilon N\chi t\hat{M}_\mathrm{qubits}}{2}\right)\rangle = \sum_{q=0}^{M}\cos^{2}\left(\frac{q\varepsilon N\chi t}{2}\right) deg(q)p(q) \\ 
\nonumber
&= \sum_{q=0}^{M}\cos^{2}\left(\frac{q\varepsilon\chi Nt}{2}\right)\binom{M}{q}p_{1}^{q}(1-p_{1})^{M-q}, \\
&= \frac{1}{\left(1+e^{-\varepsilon/T}\right)^{M}}\sum_{q=0}^{M}\binom{M}{q}\cos^{2}\left(\frac{q\varepsilon N\chi t}{2}\right)e^{-q\varepsilon/T}.
\end{align}
Note that in principle, the sum would be over $2^{M}$ different possible computational basis states, but bases with the same Hamming weight (defined as the total number of ones in their binary representation) with value q have degenerate probability value $p(q)=p_1^q(1-p_1)^q$ in the density matrix $\opi{\rho}_T$, as well as the same eigenvalue $\cos^{2}\left(\frac{q\varepsilon N\chi t}{2}\right)$ for the operator $\cos^{2}\left(\frac{\varepsilon N\chi t\hat{M}_\mathrm{qubits}}{2}\right)$ (recalling that the computational basis diagonalizes both $\opi{\rho}_T$ and $\hat{M}_\mathrm{qubits}$). Thus, instead of taking the sum over $2^M$ states, we take it over the possible Hamming weights $q$ from $0$ to $M$, and multiply them by the corresponding degeneracies given by $deg(q)=\binom{M}{q}$.

\section{QFI Calculation}
\label{app:QFI}

Let us rewrite $p_0$ using the complex exponential \(\cos(qN_\mathrm{eff}) = \frac{e^{iqN_\mathrm{eff}} + e^{-iqN_\mathrm{eff}}}{2}\), with \(N_\mathrm{eff} = \frac{\varepsilon N\chi t}{2}\):
\begin{eqnarray}
p_{0}&=& \frac{1}{4}\sum_{q=0}^{M}\binom{M}{q}\frac{e^{q[-\varepsilon/T+2iN_\mathrm{eff}]}}{(1+e^{-\varepsilon/T})^{M}} \\
&+& \frac{1}{2}\sum_{q=0}^{M}\binom{M}{q}\frac{e^{-q\varepsilon/T}}{(1+e^{-\varepsilon/T})^{M}} \\
&+& \frac{1}{4}\sum_{q=0}^{M}\binom{M}{q}\frac{e^{q[-\varepsilon/T-2iN_\mathrm{eff}]}}{(1+e^{-\varepsilon/T})^{M}}.
\end{eqnarray}
Recognizing the binomial expansion as follows:
\begin{align}
& \sum_{q=0}^{M}\binom{M}{q}\frac{e^{q[-\varepsilon/T\pm 2iN_\mathrm{eff}]}}{(1+e^{-\varepsilon/T})^{M}} = \frac{\left(1+e^{[-\varepsilon/T\pm 2iN_\mathrm{eff}]}\right)^{M}}{(1+e^{-\varepsilon/T})^{M}}, \\
& \sum_{q=0}^{M}\binom{M}{q}\frac{e^{-q\varepsilon/T}}{(1+e^{-\varepsilon/T})^{M}} = \frac{(1+e^{-\varepsilon/T})^{M}}{(1+e^{-\varepsilon/T})^{M}} = 1,
\end{align}
we can write
\begin{equation}
p_{0} = \frac{1}{4}\left[\frac{\left(1+e^{[-\varepsilon/T+2iN_\mathrm{eff}]}\right)^{M} + \left(1+e^{[-\varepsilon/T-2iN_\mathrm{eff}]}\right)^{M}}{(1+e^{-\varepsilon/T})^{M}}\right] + \frac{1}{2},
\end{equation}
or
\begin{equation}
p_{0} = \frac{1}{2}\frac{{\text{Re}}\left[\left(1+e^{[-\varepsilon/T+2iN_\mathrm{eff}]}\right)^{M}\right]}{(1+e^{-\varepsilon/T})^{M}} + \frac{1}{2}.
\end{equation}
Now, using De Moivre's formula \(z^{M} = |z|^{M}(\cos(M\theta) + i\sin(M\theta))\):
\begin{align}
p_{0} = \frac{1}{2}\left[\frac{\sqrt{1+2e^{-\varepsilon/T}\cos(2N_\mathrm{eff})+e^{-2\varepsilon/T}}}{1+e^{-\varepsilon/T}}\right]^{M}\cos(M\theta) + \frac{1}{2}, \\
\theta = \arcsin\left(\frac{e^{-\varepsilon/T}\sin(2N_\mathrm{eff})}{\sqrt{1+2e^{-\varepsilon/T}\cos(2N_\mathrm{eff})+e^{-2\varepsilon/T}}}\right).
\end{align}
Note that, as defined, \(\theta = \theta(T,N_\mathrm{eff})\) is independent of the number of two-level atoms \(M\). Let us now define
\begin{equation}
\alpha(T,N_\mathrm{eff}) = \frac{\sqrt{1+2e^{-\varepsilon/T}\cos(2N_\mathrm{eff})+e^{-2\varepsilon/T}}}{1+e^{-\varepsilon/T}},
\end{equation}
where we easily check that \(\alpha(T,N_\mathrm{eff}) \leq 1\).
Thus, using this compact notation, we can write 
\begin{equation}
p_{0} = \frac{1}{2} \alpha^{M} \cos(M\theta) + \frac{1}{2},
\end{equation}
and, as for its derivative:
\begin{equation}
\frac{dp_{0}}{dT} = \frac{1}{2} M \alpha^{M} g \cos(M\theta + \phi),   
\end{equation}
with
\begin{equation}
g = \sqrt{C^{2} + D^{2}}, \quad C = \frac{1}{\alpha} \frac{d\alpha}{dT}, \quad D = -\frac{d\theta}{dT},    
\end{equation}
and
\begin{equation}
\Phi = \arctan\left(-\frac{D}{C}\right).
\end{equation}

We can now write for the QFI, since \( p_{0} = 1 - p_{1} \):
\begin{equation}
Q(T) = \frac{1}{p_{0}(1-p_{0})} \left( \frac{dp_{0}}{dT} \right)^{2}.    
\end{equation}
Replacing \( p_{0} \) and $\frac{dp_{0}}{dT}$ we find:
\begin{equation}
Q(T)= \frac{M^{2} \alpha^{2M} g^{2} \cos^{2}(M\theta + \Phi)}{\left[1 - \alpha^{2M} \cos^{2}(M\theta)  \right]}.    
\end{equation}
The function $\alpha$ is, in particular, always positive and less than or equal to 1. In addition, for $M = N = 1$ we recover the QFI for a single two-level system probed by an MZ interferometer. Two cases are of interest in the above equation: $\alpha = 1$ and $\alpha \neq 1$. 

\subsection{Case 1: $\alpha \neq 1$}

In this case, the maximum of the QFI is obtained by differentiating $Q(T)$ and solving for $N_\mathrm{eff}$ and $M$. Note that this procedure makes $\alpha$ dependent on $N$ and $M$. By varying either the excitation of the NOON state or the number of atoms in the sample, or both, there will be a new value of $\alpha$ that maximizes the QFI.

It is to be noted that fixing the value of $\alpha$ in the QFI formula, implies having the product of a polynomial $M$ by a function $\alpha^{2M}$ that decays exponentially with $M$. Thus, it is expected that there will be growth for some values of $M$, followed by a subsequent decrease. However, the precise maximization and the obtaining value of $\alpha$ for each $N$ and $M$ leads to a linear growth with $M$, revealing the standard behavior for QFI, thus not exceeding the Heisenberg limit.

\subsection{Case 2: \(\alpha = 1\)}

This case occurs when \(\cos(2N_\mathrm{eff}) = 1\), implicating \(\sin(2N_\mathrm{eff}) = 0\) and $N_\mathrm{eff}=k\pi$. In this situation, both the denominator and the numerator in the equation for $Q$ cancel each other out, resulting in an indeterminacy of type $0/0$. To analyze this indeterminacy, it will be necessary to consider the exact forms of $C, D$, and their derivatives:
\begin{align}
C &= \frac{1}{\alpha} \frac{\partial\alpha}{\partial T} =\frac{-1}{T^2} \frac{2\varepsilon e^{-\varepsilon/T} \sin^{2}(N_\mathrm{eff})}{1 + 2e^{-\varepsilon/T} \cos(2N_\mathrm{eff}) + e^{-2\varepsilon/T}} \left[\frac{1 - e^{-\varepsilon/T}}{1 + e^{-\varepsilon/T}}\right], \\
D &= -\frac{\partial\theta}{\partial T} =\frac{-1}{T^2} \frac{2\varepsilon e^{-\varepsilon/T} \sin(N_\mathrm{eff}) \cos(N_\mathrm{eff})}{1 + 2e^{-\varepsilon/T} \cos(2N_\mathrm{eff}) + e^{-2\varepsilon/T}}.
\end{align}
The derivatives are:
\begin{align}
\frac{\partial C}{\partial N_\mathrm{eff}} &=-\frac{1}{T^2} \frac{2\varepsilon e^{-\varepsilon/T} \sin(2N_\mathrm{eff}) (1 - e^{-2\varepsilon/T})}{(1 + 2e^{-\varepsilon/T} \cos(2N_\mathrm{eff}) + e^{-2\varepsilon/T})^{2}}, \\
\frac{\partial D}{\partial N_\mathrm{eff}} &= -\frac{1}{T^2}\frac{2\varepsilon e^{-\varepsilon/T} (\cos(2N_\mathrm{eff}) + 2e^{-\varepsilon/T} + e^{-2\varepsilon/T} \cos(2N_\mathrm{eff}))}{(1 + 2e^{-\varepsilon/T} \cos(2N_\mathrm{eff}) + e^{-2\varepsilon/T})^{2}}.
\end{align}
We search for the following product of limits:
\begin{equation}
\lim_{N_\mathrm{eff}\to k\pi}Q(T) = L_{1} L_{2} L_{3}.
\end{equation}
The first limit is straightforward::
\begin{equation}
L_{1} = \lim_{N_\mathrm{eff}\to k\pi} \alpha^{2M} = 1.
\end{equation}
The second limit evaluates to zero:
\begin{equation}
L_{2} = \lim_{N_\mathrm{eff}\to k\pi} \cos^{2}\left(M\theta + \Phi\right) = \cos^{2}\left(M\lim_{N_\mathrm{eff}\to k\pi} \theta + \lim_{N_\mathrm{eff}\to k\pi} \Phi\right) = 0.
\end{equation}
In fact, since \(\lim_{N_\mathrm{eff}\to k\pi} \theta = \arcsin(0)\),
\begin{equation}
\lim_{N_\mathrm{eff}\to k\pi} \frac{D}{C} = \frac{\cos(k\pi)}{\sin(k\pi)} \left[\frac{1 - e^{-\varepsilon/T}}{1 + e^{-\varepsilon/T}}\right] \to \infty,
\end{equation}
and therefore:
\begin{equation}
\lim_{N_\mathrm{eff}\to k\pi} \Phi = \arctan\left(-\infty\right) = -\frac{\pi}{2},
\end{equation}

Finally, the third $L_{3}$ limit is an indetermination of the kind $0/0$, which can be obtained by L'Hôpital's rule:
\begin{align} \nonumber
\lim_{N_\mathrm{eff}\to k\pi}\frac{C^{2}+D^{2}}{1-\alpha^{2M}}&=\lim_{N_\mathrm{eff}\to k\pi}\frac{2C\frac{\partial{C}}{\partial{N_\mathrm{eff}}}+2D\frac{\partial{D}}{\partial{N_\mathrm{eff}}}}{-2M\alpha^{2M-1}\frac{\partial{\alpha}}{\partial{N_\mathrm{eff}}}} \\
& =\frac{1}{T^2}\frac{2\varepsilon e^{-\varepsilon/T}}{2M(1+e^{-\varepsilon/T})^{2}}.
\end{align}
Therefore, $L_{3}$ exists and is finite: The QFI at the point $x=k\pi$, $k$ integer, results in a product of 0 by a finite number, and thus is zero. In other words, the only point that results in $\alpha=1$ does not provide a growth with $M^2$, being actually a minimum of the QFI.  

Another important point to be noted is that the excitation in the nonlinear BS, embedded in the NOON state, only appears in the argument of the trigonometric functions as \(N_\mathrm{eff} = \frac{\varepsilon N\chi t}{2}\), and therefore is only relevant to adjust the time at which the maximum of the QFI occurs. This maximum, as we can see, depends on the quantity $M$ of atoms in the thermal sample. Therefore, contrary to previous claims, the temperature measurement does not scale indiscriminately with $N^2$. In fact, there are several maxima and minima, and increasing $N$ can lead from a maxima to a minima, depending on how it is done. Given these facts the conjoint use of $M$ two-level systems and nonlinear beam splitters does not enable surpassing the Heisenberg limit using the Mach-Zehnder interferometer.

\begin{figure}[t]
\centering
 \includegraphics[width=\linewidth]{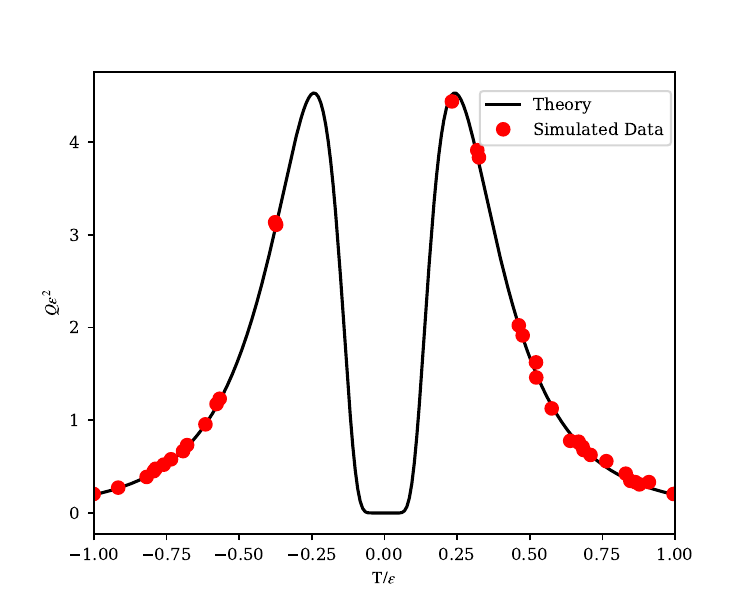}
 \caption{Quantum Fisher information (QFI) as a function of temperature $T$ for $M=N=1$.}\label{fig_comp_1}
\end{figure}

Note that for $M = N = 1$, the general result coincides with that of the simple Mach-Zehnder (MZ) interferometer. In Fig. \ref{fig_comp_1}, we show the theoretical curve and the simulation performed in Qiskit. Note that both agree perfectly.

\subsection{Dependence of QFI with key parameters.}

\begin{figure*}[t]
\centering
\includegraphics[width=0.9\linewidth]{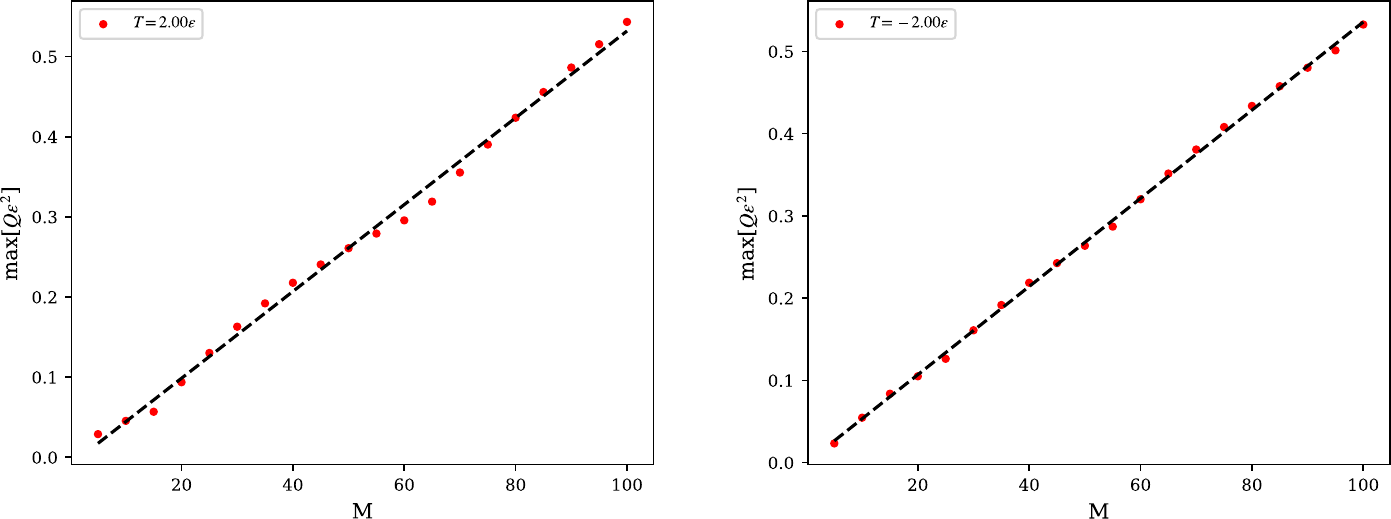}
\caption{Maximum QFI as a function of number of two-level atoms $M$ for $T=2\varepsilon$ (on the left) and $T = -2\varepsilon$ (on the right); we considered $\varepsilon=1.0$.}
\label{QFI fit}
\end{figure*}

In Fig.\ref{QFI fit}, we plot the behavior of the QFI calculated for $N_\mathrm{eff}$ that maximizes it, for different temperatures, as a function of the number of two-level atoms $M$. As seen from the figure, a clear linear scaling with respect to $M$ is shown.

\begin{figure*}[t]
\centering
\includegraphics[width=0.9\linewidth]{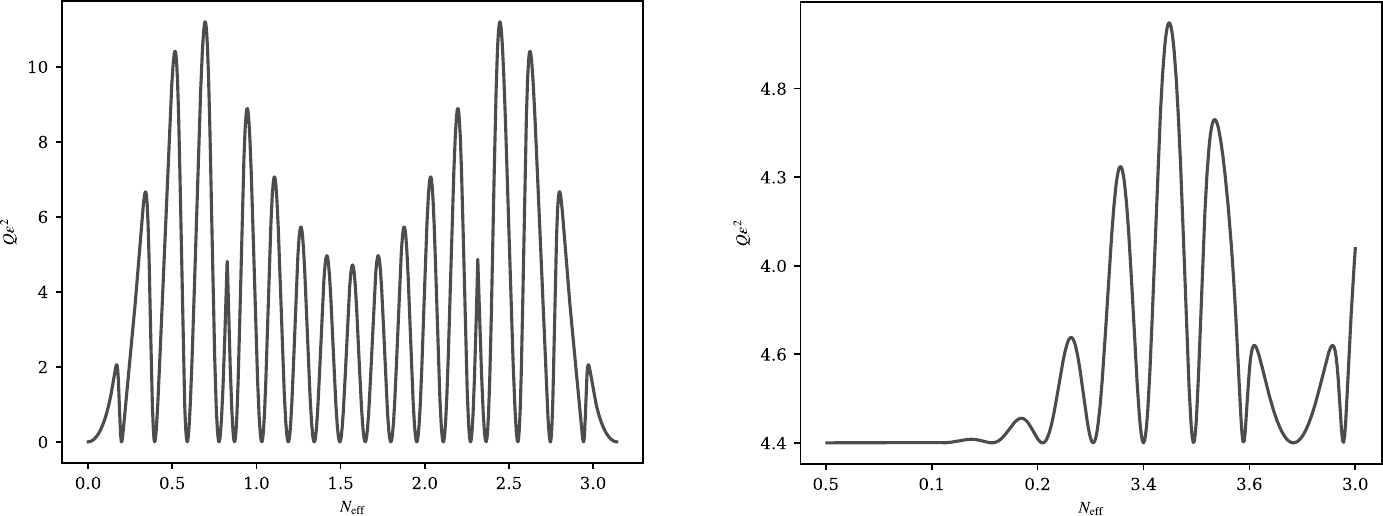}
\caption{QFI as a function of $N_\mathrm{eff} = \varepsilon N\chi t/{2}$ for different fixed values of $M, T$ and $\varepsilon$. Very close maxima and minima and abrupt changes can be observed, making the choice of optimal parameters very challenging.}
\label{fig: qfi v x}
\end{figure*}

As mentioned above, as $M$ increases, to maximize $Q(T)$ it is necessary to choose the value of $N_\mathrm{eff}=\frac{\varepsilon N\chi t}{2}$ very precisely. This is illustrated in Fig.\ref{fig: qfi v x}, which shows that the QFI changes abruptly as $N_\mathrm{eff}$ changes. Note that QFI is practically zero for $N_\mathrm{eff}<2.8$ when $T=2\varepsilon$.

\begin{figure}[t]
\centering \includegraphics[width=\linewidth]{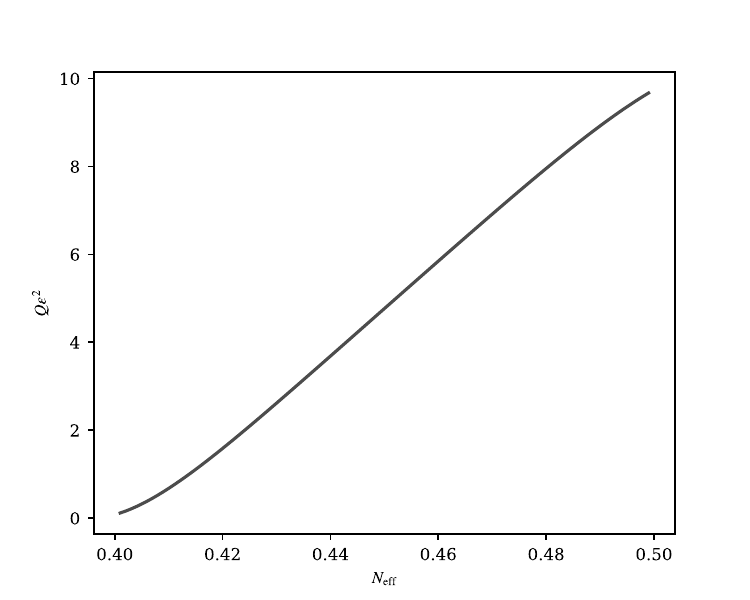}
\caption{QFI as a function of $N_\mathrm{eff}$ for $M=9,T=-0.4\varepsilon, \varepsilon=1.0$. Heisenberg limit cannot be observed.}\label{fig: linear}
\end{figure}

A closer observation of the variation between a minimum and the global maximum for $M=9, T=-0.4\varepsilon, \varepsilon=1.0$, as seen in Fig. \ref{fig: linear}, shows an abrupt growth with nearly constant rate (linear), so there is no Heisenberg limit growth.

\begin{figure*}[t]
\centering
\includegraphics[width=0.9\linewidth]{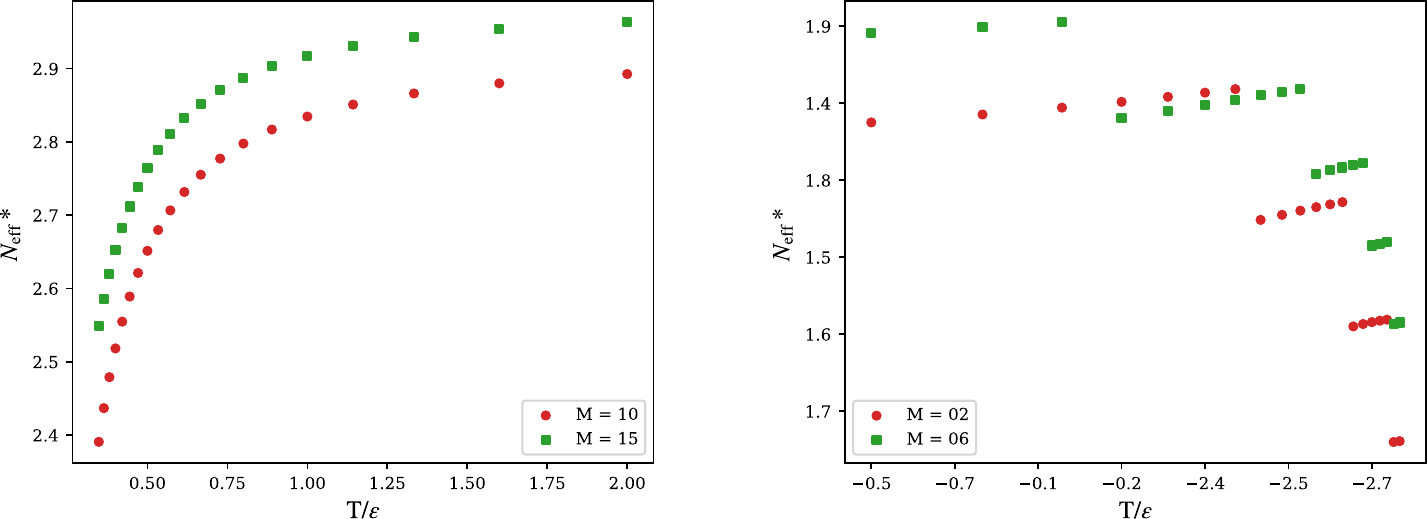}
\caption{The parameter $N_\mathrm{eff} = \frac{\varepsilon M \chi t}{2}$ that maximizes the QFI versus the number of two-level atoms in the thermal sample (with $\varepsilon=1.0$). Note that $N_\mathrm{eff}$ varies significantly with positive temperatures (left), while at effective negative temperatures (right)  decreases very little, indicating that a wide range of temperatures can be measured without the need to adjust $N_\mathrm{eff}$ for each temperature value.}
\label{chi v n}
\end{figure*}

Here it is important to highlight another factor. The behavior of the parameter $N_\mathrm{eff} = \frac{\varepsilon M \chi t}{2}$ that maximizes the QFI is not symmetric for positive and effective negative temperatures. As shown in Fig.\ref{chi v n}, while $N_\mathrm{eff}$ varies significantly with positive $T$ (on the left), for effective negative $T$ (on the right) it changes very little, indicating that a wide range of temperatures can be estimated optimally by keeping $N_\mathrm{eff}$ constant.

\end{appendix}

\section*{Declarations}

\subsection*{Availability of data and materials}
The datasets generated and analyzed during the current study are available from the corresponding author upon reasonable request.

\subsection*{Competing interests}
The authors declare that they have no competing interests.

\subsection*{Funding}
This work was supported by  support from the Brazilian agency, CAPES (Financial code 001), CNPq, and FAPEG. This work was performed as part of the Brazilian National Institute of Science and Technology (INCT) for Quantum Information Grant No. 465469/2014-0. V.M. thanks support from the National Natural Science Foundation of China Grants No. 12374482 and W2432005. NGA acknowledges Grant FAPESP 2024/21707-0.

\subsection*{Abbreviations}
QFI, quantum Fisher information;  MZI, Mach–Zehnder interferometer; QET, quantum estimation theory; CFI, classical Fisher information; POVM, positive operator-valued measure; BS, Beamsplitter ; NISQ, Noisy Intermediate-Scale Quantum.

\subsection*{Authors' contributions}
D.Y.A. and L.F.R.M. made equal contributions to the theoretical development, data analysis, and implementation of experimental simulations using IBMQ systems. G.G.D. supported the theoretical work and prepared the figures. N.G.A., G.D.M.N., and V.M. collaboratively proposed and supervised the project and contributed to finalizing the manuscript. All authors reviewed and approved the final version.

\section*{Acknowledgments}

Authors acknowledge IBM for their IBM\_brisbane quantum processor used for this study.

\bibliography{References}

\end{document}